\documentclass{ieeeaccess}
\usepackage{cite}
\usepackage{textcomp}

\usepackage{graphicx}
\usepackage{amsmath,amssymb,amsfonts}
\usepackage{multirow}
\usepackage{psfrag}
\usepackage{subfigure}
\usepackage{color}
\usepackage{soul}
\usepackage{xcolor}
\usepackage{lipsum}
\usepackage[ruled,vlined,linesnumbered]{algorithm2e}
\usepackage{mathrsfs}
\usepackage{pbox}

\def\BibTeX{{\rm B\kern-.05em{\sc i\kern-.025em b}\kern-.08em
    T\kern-.1667em\lower.7ex\hbox{E}\kern-.125emX}}

\makeatletter
\def\hlinewd#1{%
\noalign{\ifnum0=`}\fi\hrule \@height #1 %
\futurelet\reserved@a\@xhline}
\makeatother

\graphicspath{{Figures/}}

\begin{document}
\history{Date of publication xxxx 00, 0000, date of current version xxxx 00, 0000.}
\doi{10.1109/ACCESS.2017.DOI}

\title{SaferCross: Enhancing Pedestrian Safety Using Embedded Sensors of Smartphone}
\author{\uppercase{Myounggyu Won}\authorrefmark{1}, \IEEEmembership{Member, IEEE},
\uppercase{Aawesh Shrestha}\authorrefmark{2}, \uppercase{Kyung-Joon Park}\authorrefmark{3}, \IEEEmembership{Member, IEEE}, and \uppercase{Yongsoon Eun}\authorrefmark{3}, \IEEEmembership{Senior Member, IEEE}}
\address[1]{Department of Computer Science, University of Memphis, Memphis, TN 38152, USA}
\address[2]{Department of Electrical Engineering and Computer Science, South Dakota State University, Brookings, SD 57006, USA}
\address[3]{Information and Communication Engineering Department, Daegu Gyeongbuk Institute of Science and Technology (DGIST), Daegu, South Korea}
\tfootnote{This research was supported in part by the Competitive Research Grant Program (CRGP) of South Dakota Board of Regents (SDBoR), and in part by Global Research Laboratory Program (2013K1A1A2A02078326) through NRF, and DGIST Research and Development Program (CPS Global Center) funded by the Ministry of Science, ICT \& Future Planning of South Korea.}

\markboth
{Author \headeretal: Preparation of Papers for IEEE TRANSACTIONS and JOURNALS}
{Author \headeretal: Preparation of Papers for IEEE TRANSACTIONS and JOURNALS}

\corresp{Corresponding authors: Kyung-Joon Park (e-mail: kjp@dgist.ac.kr) and Yongsoon Eun (email: yeun@dgist.ac.kr).}

\begin{abstract}
The number of pedestrian accidents continues to keep climbing. Distraction from smartphone is one of the biggest causes for pedestrian fatalities. In this paper, we develop $\mathsf{SaferCross}$, a mobile system based on the embedded sensors of smartphone to improve pedestrian safety by preventing distraction from smartphone. $\mathsf{SaferCross}$ adopts a holistic approach by identifying and developing essential system components that are missing in existing systems and integrating the system components into a ``fully-functioning'' mobile system for pedestrian safety. Specifically, we create algorithms for improving the accuracy and energy efficiency of pedestrian positioning, effectiveness of phone activity detection, and real-time risk assessment. We demonstrate that $\mathsf{SaferCross}$, through systematic integration of the developed algorithms, performs situation awareness effectively and provides a timely warning to the pedestrian based on the information obtained from smartphone sensors and Direct Wi-Fi-based peer-to-peer communication with approaching cars. Extensive experiments are conducted in a department parking lot for both component-level and integrated testing. The results demonstrate that the energy efficiency and positioning accuracy of $\mathsf{SaferCross}$ are improved by 52\% and 72\% on average compared with existing solutions with missing support for positioning accuracy and energy efficiency, and the phone-viewing event detection accuracy is over 90\%. The integrated test results show that $\mathsf{SaferCross}$ alerts the pedestrian timely with an average error of 1.6sec in comparison with the ground truth data, which can be easily compensated by configuring the system to fire an alert message a couple of seconds early.
\end{abstract}

\begin{keywords}
Mobile Computing, Pedestrian Safety, Wi-Fi Direct
\end{keywords}

\titlepgskip=-15pt

\maketitle

\section{Introduction}
\label{sec:introduction}

The number of pedestrian accidents continues to keep climbing. In 2018, 6,283 pedestrians were killed which accounted for an increase of 3\% compared with pedestrian fatalities in 2017, the highest number of pedestrian fatalities since 1990~\cite{pedestrians}.

Many sources point out that smartphone distraction is one of the major causes for pedestrian fatalities~\cite{ped_stat1}\cite{ped_stat2}\cite{ped_stat3}. Many pedestrians use their mobile phones while walking on sidewalks and crossing the street{~\cite{basch2015pedestrian}}{~\cite{thompson2013impact}}. A recent study shows that more than a third of pedestrians use their mobile phones while crossing streets{~\cite{cbs}}, and 16\% of pedestrian accidents were caused by distraction due to phone use{~\cite{ped_stat4}}. Another study shows that 85\% have seen distracted pedestrians, and 26\% of the respondents were actually involved with distracted-walking accidents{~\cite{lin2017impact}}. According to the report from US Consumer Product Safety Commission, the percentage of pedestrian injuries involving smartphones are increasing steadily{~\cite{nasar2013pedestrian}}. Interestingly, the significant rise in the pedestrian injuries started in 2009, and this is exactly when smartphones started to take hold{~\cite{androidpit2}}. These distracted pedestrians are even called ``smartphone zombies'' in recent scientific publications to stress the seriousness of the problem{~\cite{duke2017smartphone}}.

In this paper, we aim to develop a mobile system that enhances pedestrian safety by preventing distracted phone use. Numerous approaches have been designed and deployed to protect distracted pedestrians, \emph{e.g.,} using a signage{~\cite{lookupny}}, and LED spotlights{~\cite{ledspot}} as shown in Fig.{~\ref{fig:ex_systems}}. However, painting road signs and installing LED lights at every crosswalk not only involves huge cost, but it can even be distraction to drivers especially at night. With a myriad of embedded sensors such as accelerometers, cameras, microphones, and GPS, smartphones have opened the new opportunities for various applications{~\cite{fahim2017alert}}; especially, those sensors can be used to improve pedestrian safety by directly alerting pedestrians. The camera of smartphone is used to detect approaching vehicles posing danger to pedestrians~\cite{wang2012walksafe}. Some mobile systems utilize communication between cars and pedestrians. For example, Wu \emph{et al.} adopt the Dedicated Short Range Communication (DSRC) to enable the vehicle to pedestrian communication~\cite{wu2014cars}, and Lin \emph{et al.} utilize the cellular network~\cite{lin2016psafety} to perform risk analysis and alert the pedestrian accordingly.

Although existing solutions contribute to improving pedestrian safety, most solutions are focused on a certain aspect of a mobile system for pedestrian safety such as detection of approaching cars, communication between cars and pedestrians. However, a number of essential system components are still missing to build a complete mobile system for pedestrian safety. In this paper, \emph{we aim to develop a ``fully-functioning'' mobile system for pedestrian safety by developing these critical system components for precise pedestrian positioning, energy efficiency, accurate phone activity detection, and effective risk assessment.} To this end, we present $\mathsf{SaferCross}$ , a mobile system for pedestrian safety based on the embedded sensors and WiFi Direct of smartphone~\cite{camps2013device}. $\mathsf{SaferCross}$ effectively senses approaching cars, performs real-time risk assessment, and alerts drivers and pedestrians in a timely and energy-efficient manner without requiring any modifications to the host mobile system. The key contributions of $\mathsf{SaferCross}$ are development of the essential system components that can be adopted for developing mobile systems for pedestrian safety, ultimately to spark the mobile computing research focused on protecting pedestrians from accidents. We note that it is ideal to force pedestrians not to use their phones while walking. At the same time, however, we admit that there are still a lot of pedestrians who are tempted to use their phones. In fact, a huge number of pedestrian involved accidents happen every year. As such, we argue that $\mathsf{SaferCross}$ is developed as an assistive technology for those distracted pedestrians and is not intended to encourage blind reliance on the technology.

\Figure[t](topskip=0pt, botskip=0pt, midskip=0pt)[width=.99\columnwidth]{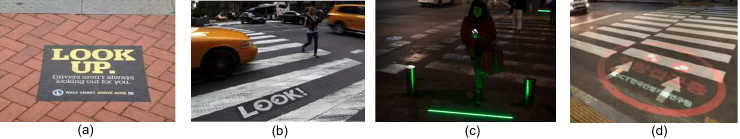}
{(a) Signage in Delaware, (b) Signage in NYC, (c) LED stoplight in Israel, (d) Ground flashlight in Seoul. Numerous pedestrian safety systems have been developed.\label{fig:ex_systems}}

$\mathsf{SaferCross}$ is built upon fundamental technologies focused on enhancing the effectiveness of pedestrian localization, energy efficiency of smartphone, phone activity detection, and situation awareness. More specifically, the pedestrian localization module of $\mathsf{SaferCross}$ is specifically designed to accurately localize slow-moving pedestrians. This module is particularly useful for localizing pedestrians in urban areas with skyscrapers where accurate positioning based on GPS is very challenging. Based on the observation that pedestrians walk along a sidewalk, a map matching algorithm is adopted and customized for accurately localizing slow-moving pedestrians. To mitigate the impact of significant power consumption of the GPS module and improve the energy efficiency, a dynamic approach is proposed to activate the GPS module adaptively depending on the estimated time that the user is expected to be geographically close to a nearby crosswalk. Additionally, a novel algorithm is designed to detect the user activities effectively so that the system is activated at the exactly right time. The communication module of $\mathsf{SaferCross}$ is built based on the WiFi Direct technology. A new technique based on opportunistic overhearing of WiFi Direct messages is developed to address the challenge of allowing for n-to-n communication for WiFi Direct. Finally, a collision probability model is designed based on real-world data collected via WiFi Direct to effectively perform risk assessment. $\mathsf{SaferCross}$ systematically structures these novel system components to build the first fully functioning mobile system for pedestrian safety.

Extensive experiments were conducted in which both the module-level test and the integrated test were performed to  evaluate the effectiveness of the proposed system modules individually and the system performance as a whole. The results demonstrate that the localization accuracy was significantly improved by up to 72\%; the user phone viewing event was accurately detected with the accuracy of over 90\%; and the energy consumption was reduced by 50.2\%. An integrated test was performed to evaluate the effectiveness of the interplay of the individual system components. The results show that all modules effectively cooperate to provide an alert message to the user in a timely manner with an error of 1.6sec on average in comparison with the ground truth data. Such a small error can be easily compensated by configuring the system to fire an alert message a couple of seconds early. The contributions of this paper are summarized as follows.

\begin{itemize}
\item A Hidden Markov Model-based map matching algorithm is designed for accurately localizing pedestrians.
\item An adaptive algorithm is developed to improve the energy efficiency of a mobile system for pedestrian safety.
\item An effective algorithm is created to detect the pedestrian phone viewing event accurately.
\item A novel approach is developed to perform risk assessment effectively based on Wi-Fi Direct-based communication between cars and pedestrians.
\item Experiments are performed in a parking lot to demonstrate the effectiveness of individual system components and the proposed mobile system as a whole.
\end{itemize}

This paper is organized as follows. Section{~\ref{sec:related_work}} presents a literature review on related approaches designed for improving pedestrian safety. In Section~\ref{sec:system_design}, we describe an overview of the proposed app followed by the details of each system component. The performance of the proposed app is evaluated in Section~\ref{sec:experimental_results}. We then conclude in Section~\ref{sec:conclusion}.

\section{Related Work}
\label{sec:related_work}

Wang \emph{et al.} develop an app that uses the rear camera of a phone to monitor approaching vehicles to alert the pedestrian~\cite{wang2012walksafe}. A machine-learning-based image processing algorithm is designed to capture approaching cars for pedestrian safety assessment. This approach, however, raises the privacy issue as it takes photos of cars without acquiring permission of drivers. Additionally, the energy efficiency is another problem as this system is based on continuously executing image processing algorithms which consume much energy.

A cellular network is used to enable car-to-pedestrian communication~\cite{lin2016psafety}. However, this approach based on a cellular network not only incurs high cost but also results in non-negligible message delay compared with the direct peer to peer communication. Especially, even a small message delay is critical in mobile systems for pedestrian safety. Dedicated Short Range Communication (DSRC) is a wireless communication standard specifically designed for vehicle-to-vehicle communication (V2V). Researchers utilize DSRC as a means to enable vehicle-to-person (V2P) communication for pedestrian safety~\cite{wu2014cars}. However, implementing DSRC on a phone requires significant modifications to the host system firmware, and extra device support is needed to operate DSRC on vehicles.

%For many pedestrian safety systems, position information of pedestrians and cars is critical in determining the dangerous situation. Although a rich literature exists on this topic ~\cite{jain2014limits}\cite{datta2014towards}\cite{lin2016psafety}, there is few in-depth discussion on the positioning accuracy,

%Position information of a pedestrian and a car is used to determine the dangerous situation~\cite{jain2014limits}\cite{datta2014towards}\cite{lin2016psafety}. Although some approaches use a sector rather a location by formulating a sector overlapping problem to see if a collision would occur to compensate for the GPS location error~\cite{lin2016psafety}, no in-depth discussion on when, how, and to whom an alert message is sent is missing.

Specialized hardware is designed to enhance pedestrian safety. For example, sensors are adhered to the pedestrians' shoes to detect whether the pedestrian is crossing at a crosswalk~\cite{jain2015lookup}. Those sensors are used to calculate the slope between the sidewalk and the roadway as an indicator to find whether the pedestrian is about to cross the street. Another example is to exploit an electronic transponder that is attached to the pedestrian's body to determine whether the pedestrian is visible or not~\cite{fackelmeier2008dual}. However, typically asking the users to attach these types of specialized hardware is not easy, preventing widespread adoption of such technology.

WiFi has been actively considered as an appropriate alternative technology to enable vehicle to pedestrian communication for pedestrian safety~\cite{anaya2014vehicle}\cite{dhondge2014wifihonk}\cite{ho2016wisafe}. In particular, WiHonk is quite similar to the \textsf{Communication} module of our work~\cite{dhondge2014wifihonk}. However, WiHonk is based on the modification of the beacon frame of IEEE 802.11 which requires the root privilege that makes it difficult for common use. Additionally, no details are provided regarding when to exchange messages with cars, potentially resulting in unnecessary network bottleneck. WiSafe is another WiFi-based pedestrian safety system which resembles our \textsf{Communication} module~\cite{ho2016wisafe}. Our work is different in that the system design involves both the driver and pedestrian while WiSafe utilizes only one-way communication from a pedestrian to cars.

%\emph{However, the most important point is that these solutions are partially developed, and as discussed in this paper, there are a lot of critical components to consider, develop and structure systematically in order to have a fully functioning mobile system for pedestrian safety.} The key contribution of $\mathsf{SafePedCross}$ is thus to build the solid foundation for further research on mobile system development for pedestrian safety.

\section{System Design}
\label{sec:system_design}

\subsection{System Overview}
\label{sec:sys_overview}

\begin{table*}[t]
\center
\label{table:notations}
\caption{Notations used in this paper.}
\begin{tabular}{ l|l }
  \hlinewd{1.5pt}
  %\multicolumn{2}{c}{Notations Used in This Paper} \\
  %\hline
  $r_i, 0 \le i \le N$ & HMM state representing a sidewalk segment \\
  $\lambda = (S,Z_{t},A,B,\pi)$ & HMM model \\
  $S=\{r_1,r_2,...,r_N\}$ & State set \\
  $Z_t=\{z_1,...z_{\omega}\}$ & Set of preceding GPS locations \\
  $\omega$ & Size of the sliding window to save $Z_t$ \\
  $A = P(Z_t | r_i), 1 \le i \le N$ & Observation probabilities \\
  $B = P(r_{j}|r_{i}), i \neq j (i,j=1...N)$ & Transition probabilities \\
  $\pi = P(Z_1|r_i), 1 \le i \le N$ & Initial state probabilities \\
  $z_{t,i}$ & Geographically closest location from $z_t$ on a sidewalk segment $r_i$ \\
  $|z_i - z_j|_{geo}$ & Geodetic distance between $z_i$ and $z_j$ \\
  $\sigma_z$ & Standard deviation of GPS measurements \\
  $|z_i - z_j|_{mov}$ & Moving distance between $z_i$ and $z_j$ \\
  $\alpha$ & Parameter for adjusting the tolerance level against location error \\
  $v_{max}$ & Maximum brisk walking speed \\
  $\{a_x,a_y,a_z\}$ & Accelerometer reading \\
  $\{\hat{a_x}, \hat{a_y}, \hat{a_z}\}$ & Filtered accelerometer reading \\
  $v_c(i)$ & Speed of approaching vehicle $i$ \\
  $m(i)$ & Vehicle mass of approaching vehicle $i$ \\
  $A(i)$ & Cross-sectional area of approaching vehicle $i$ \\
  $t_c(i)$ & Time for reaching the crossing for approaching vehicle $i$ \\
  $v_p$ & User walking speed \\
  $t_p$ & Time for the user to reach the crossing \\
  $t_{warning}$ & User warning time \\
  $t_{delay}$ & Round trip delay for a single hop 802.11 link \\
  $t_{react}$ & Driver reaction delay \\
  $t_{skid}$ & Time between applying the brakes and complete stop \\
   \hlinewd{1.5pt}
\end{tabular}
\end{table*}

$\mathsf{SaferCross}$ has two modes of operation: the driver mode and pedestrian mode. In the driver mode, $\mathsf{SaferCross}$ keeps monitoring the speed and location of the vehicle, and sends the speed and location information to the pedestrians within the communication range of WiFi Direct. In the pedestrian mode, $\mathsf{SaferCross}$ keeps track of the pedestrian location and activity to detect if the user is attempting to cross a crossing while viewing their phone. It communicates with approaching cars, \emph{i.e.,} driver phones of those cars, to obtain the speed and location information of the cars and estimates the probability of collision. Depending on the calculated probability of collision, the pedestrian is alerted. To minimize driver distraction, the driver is only alerted when the pedestrian ignores the alert message several times.

\Figure[t](topskip=0pt, botskip=0pt, midskip=0pt)[width=.75\columnwidth]{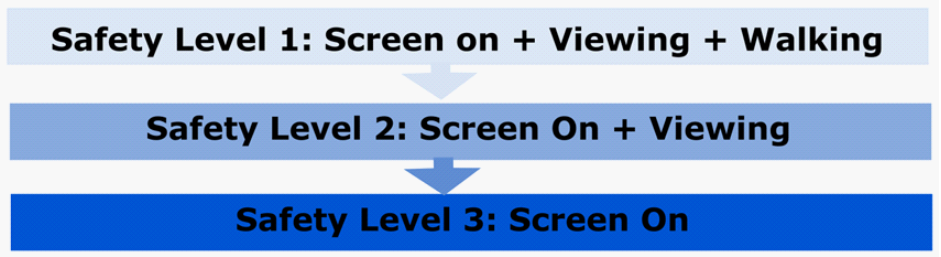}
{Different pedestrian safety levels of $\mathsf{SaferCross}$. $\mathsf{SaferCross}$ can be configured to support various pedestrian safety levels.\label{fig:safety_level}}

\Figure[t!](topskip=0pt, botskip=0pt, midskip=0pt)[width=.8\textwidth]{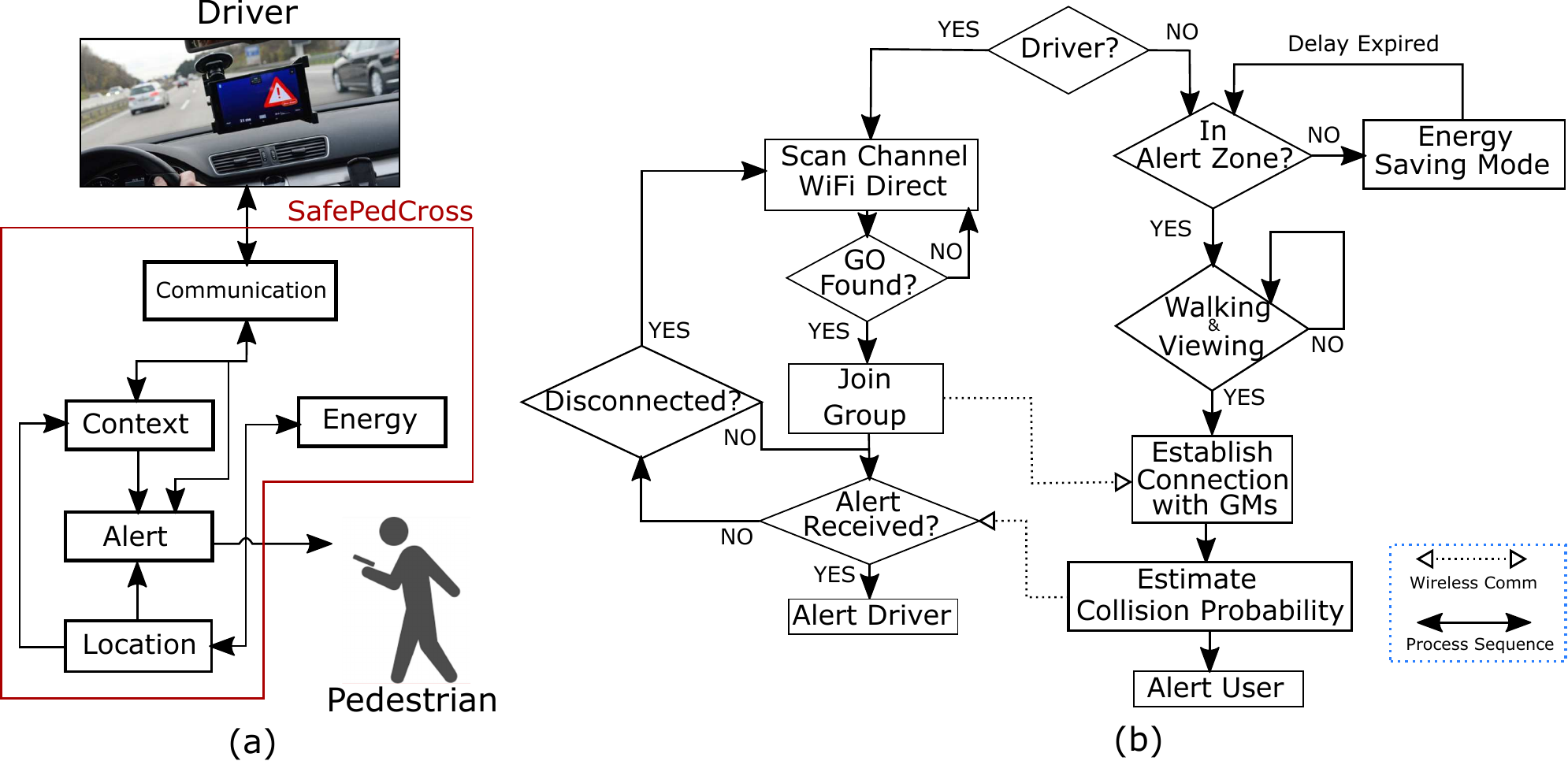}
{(a) System structure of $\mathsf{SaferCross}$; (b) Flow chart representing the operation of $\mathsf{SaferCross}$. \label{fig:structure}}

$\mathsf{SaferCross}$ also supports the stand-alone mode where it works without requiring to communicate with driver phones. In other words, it can be configured to alert the pedestrian based only on the distance to the crossing and the pedestrian's walking direction with varying safety levels (Fig.~\ref{fig:safety_level}). More specifically, the user will be alerted if he is close to the crossing and his phone screen is on (Level 3); if the phone screen is on and he is viewing the phone (Level 2); and if the phone screen is on and he is viewing the phone and he is walking toward the crossing (Level 1).

%In this paper, we address the five main questions to enable the first fully-functioning mobile P2P system for pedestrian safety.

%\begin{itemize}
%  \item How can we obtain accurate user locations using the smartphone GPS especially under urban environments?
%  \item How can we minimize power consumption of phone when we have to continually keep track of the user location?
%  \item How can we effectively detect the particular user activity of `walking while viewing phone'?
%  \item How can we model the collision probability to effectively estimate the chance of accident?
%  \item How can we ensure that the pedestrian's phone communicates effectively with the driver's phone especially when there are many cars and pedestrians participating in the communication?
%\end{itemize}

$\mathsf{SaferCross}$ consists of five main system modules, namely \textsf{Location}, \textsf{Energy}, \textsf{Context}, \textsf{Alert}, and \textsf{Communication} (Fig.~\ref{fig:structure}(a)). Specifically, the \textsf{Location} module is developed to improve the positioning accuracy of the pedestrian. The resulting pedestrian location information is distributed to other system modules. The \textsf{Energy} module is designed to save energy by adaptively controlling the operation of the GPS module. Taking the user location as input from the \textsf{Location} module, the \textsf{Context} module identifies the user activity, \emph{e.g.,} whether the user is walking, running, and viewing their phone. In particular, the module addresses the challenge of effectively detecting the `phone viewing' activity. The \textsf{Alert} module is where the collision probability is calculated. As can be seen in Fig.~\ref{fig:structure}(a), it interacts with the \textsf{Location} and \textsf{Communication} modules to obtain necessary information in calculating the collision probability. The \textsf{Alert} module then makes a decision to send an alert message to the user based on the resulting collision probability. The \textsf{Communication} module enables P2P communication between pedestrians and approaching cars using WiFi Direct.

%\begin{figure}[!htbp]
%\centering
%\includegraphics[width=.7\columnwidth]{system_overview}
%\caption {System operation of $\mathsf{SafePedCross}$.}
%\label{fig:system_overview}
%\end{figure}

To explain the operation of $\mathsf{SaferCross}$ in more detail, a flowchart (Fig.~\ref{fig:structure}(b)) is used. When the system is started, it identifies whether the user is a driver or not. For this, we adopt an existing driver phone detection algorithm~\cite{wang2013sensing}. If the user is a driver, using WiFi Direct, $\mathsf{SaferCross}$ starts to scan on the predetermined channel to be connected with pedestrians. Once it is connected, $\mathsf{SaferCross}$ sends the vehicle information to the pedestrian so that the pedestrian can calculate the collision probability. More specifically, the \emph{autonomous} mode of WiFi Direct is adopted to minimize the connection establishment time and to alert the user timely~\cite{camps2013device}.

If the user is a pedestrian, the \textsf{Location} module is activated to obtain the calibrated user location. This location information is distributed to the \textsf{Energy},  \textsf{Context}, and \textsf{Alert} modules. The \textsf{Energy} module in turn finds if the user is located within an alert zone, a region around a crossing--detailed description about the alert zone will be presented when we explain the \textsf{Energy} module in Section~\ref{sec:energy_engine}. If the user is in an alert zone, the \textsf{Context} module kicks in and detects whether the user is actually walking/running while viewing their phone. If the phone-viewing-event is detected, the \textsf{Context} module triggers the \textsf{Communication} module. And then the \textsf{Communication} module is used to create a P2P group for Wi-Fi Direct to initiate communication with cars and obtain necessary information for the \textsf{Alert} module to perform risk assessment. Notations used to explain the modules of $\mathsf{SaferCross}$ throughout this paper are summarized in Table 1.

\subsection{Improving Positioning Accuracy}
\label{sec:gps_inaccuracy}

%\begin{wrapfigure}{r}{2.0in}
%\vspace{-10pt}
%\centering
%\includegraphics[width=0.45\columnwidth]{location_experiment}
%\caption{Location errors in metropolitan area.}
%\label{fig:location_experiment}
%\vspace{-7pt}
%\end{wrapfigure}

\Figure[t](topskip=0pt, botskip=0pt, midskip=0pt)[width=.9\columnwidth]{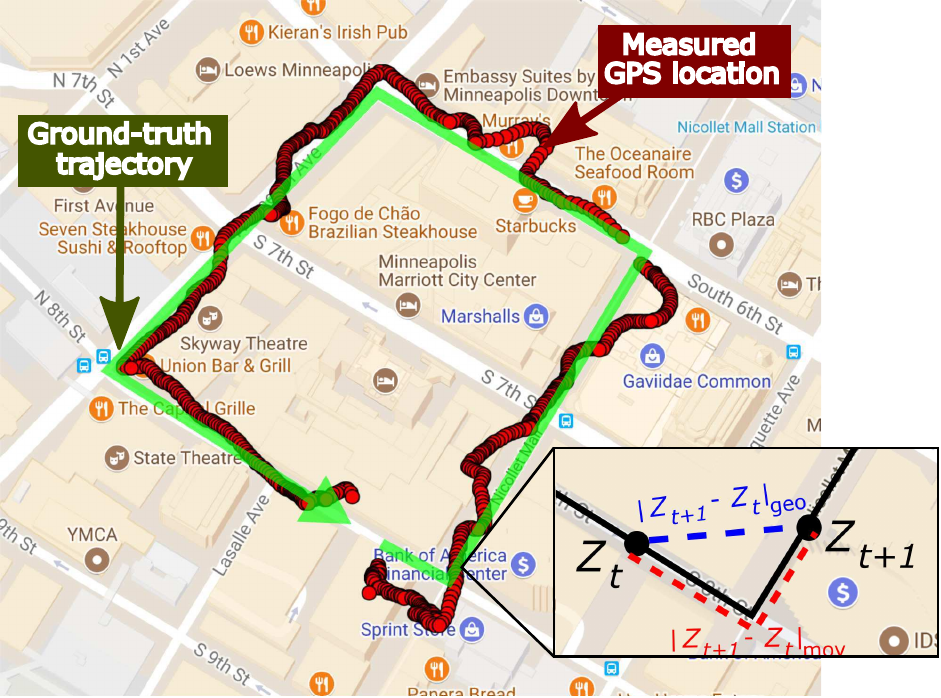}
{Location measurement in a metropolitan area. The result demonstrates significantly large location errors in a typical city environment.\label{fig:location_experiment}}

%\begin{figure}[!htbp]
%\centering
%\includegraphics[width=.6\columnwidth]{location_experiment}
%\caption {Collected GPS locations and ground-truth trajectory.}
%\label{fig:location_experiment}
%\end{figure}

%\begin{figure}[!htbp]
%\centering
%\includegraphics[width=.5\columnwidth]{location_experiment}
%\caption {Location errors in metropolitan area.}
%\label{fig:location_experiment}
%\end{figure}

Attaining high positioning accuracy of the pedestrian is crucial for $\mathsf{SaferCross}$ to estimate the collision probability accurately and alert the pedestrian timely. However, achieving precise localization using the GPS module of smartphone is a challenging problem, especially in urban canyons with significant multipath and non-line-of-sight effects. To understand the localization accuracy of the smartphone that we used in our experiments, we collected GPS locations in a metropolitan area. Fig.~\ref{fig:location_experiment} shows the collected GPS locations. The red-colored dots represent the measured GPS locations. The green arrow indicates the ground-truth trajectory. The mean location error was very large as 12.9m.

\Figure[t](topskip=0pt, botskip=0pt, midskip=0pt)[width=.9\columnwidth]{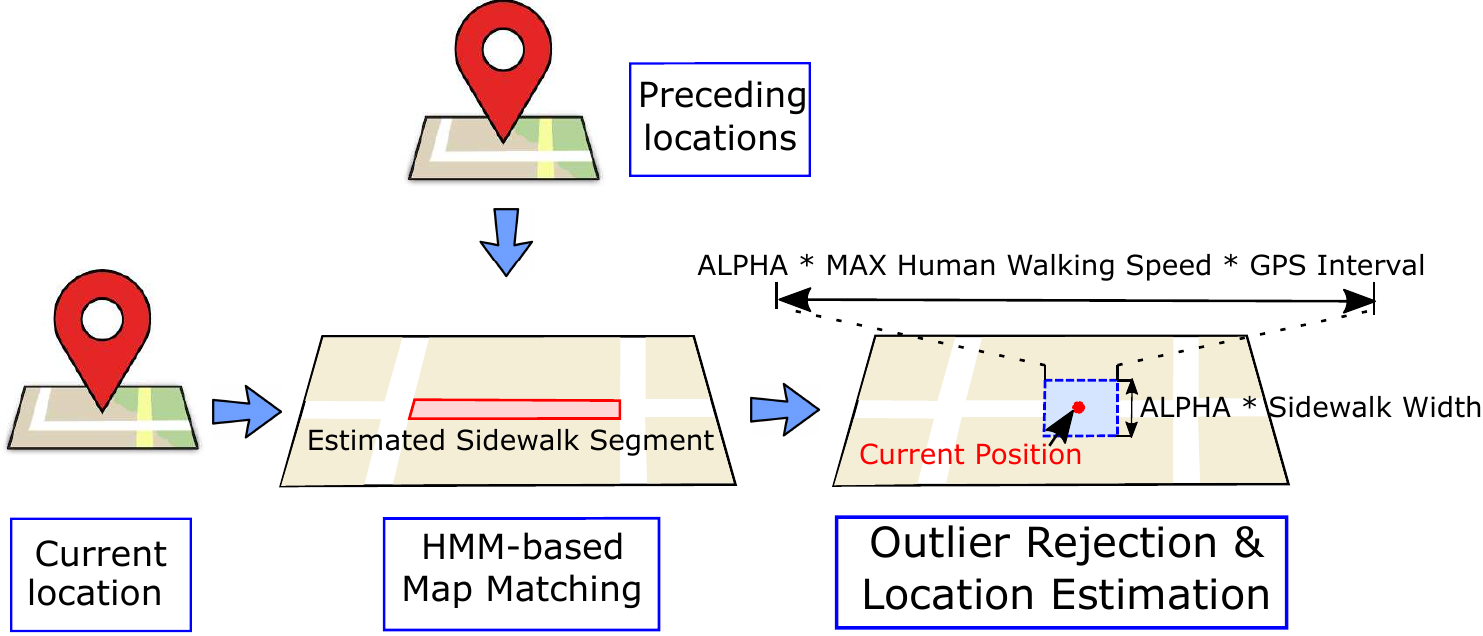}
{Overview of the \textsf{Location} module of $\mathsf{SaferCross}$. The module is developed based on a customized map matching algorithm. \label{fig:overview_positioning}}

The \textsf{Location} module of $\mathsf{SaferCross}$ is designed to improve the positioning accuracy. It finds highly erroneous GPS locations and replaces them with newly estimated locations. The \textsf{Location} module is developed based on the observation that pedestrians walk along a sidewalk, and therefore a GPS location that is geographically far from a sidewalk can be considered as an outlier. Specifically, a Hidden Markov Model-based map matching algorithm is designed to infer the current sidewalk segment using preceding user locations and to remove/replace erroneous GPS locations. In contrast to existing map matching algorithms, an unique approach is developed specifically for `slow-moving' pedestrians. Fig.~\ref{fig:overview_positioning} depicts an overview of the \textsf{Location} module. The current GPS location is provided as input to the map matching algorithm. The algorithm then estimates the current sidewalk segment based on the preceding user locations. Once the current sidewalk segment is identified, the algorithm calculates the location error and rejects or replaces the GPS location with a newly estimated location.

The \textsf{Location} module identifies the current sidewalk segment using a Hidden Markov Model (HMM). Let us define a set of states $S=\{r_1,r_2,...,r_N\}$ where each state represents a sidewalk segment with $N$ being the total number of states. Note that only the sidewalk segments in the surrounding area of the current user location are considered in finding the current sidewalk segment in order to reduce the computational overhead. Now we exploit HMM to find the most probable sidewalk segment $r_i \in S, 1 \le i \le N$ given the observation of a set of the preceding GPS locations in a sliding window examined at time $t$, which is denoted by $Z_t$. A unique aspect of the proposed map-matching algorithm based localization method compared to other map matching algorithms is that a set of preceding locations are taken into account rather than a single location to account for the low speed of a pedestrian.

More formally, a HMM is modeled as $\lambda = (S,Z_{t},A,B,\pi)$, where $S$ is the state set. $Z_t$ is an observation that is represented as a sliding window of size $\omega$ consisting of the preceding GPS locations, \emph{i.e.,} $Z_t=\{z_1,...z_{\omega}\}$, where $z_j$ is a GPS location measured at time $j$. $A$ is the observation probabilities denoted by $P(Z_t | r_i), 1 \le i \le N$. It defines the likelihood that the user is actually on sidewalk segment $r_i$. $B$ is the transition probabilities denoted by $P(r_{j}|r_{i}), i \neq j (i,j=1...N)$. It represents the likelihood of the user moving from one segment $r_{i}$ to another $r_{j}$. $\pi$ is the initial state probabilities which are defined as $P(Z_1|r_i), 1 \le i \le N$.

%\begin{figure}[!htbp]
%\centering
%\includegraphics[width=0.4\columnwidth]{moving_distance_small}
%\caption{Illustration of moving distance and geodetic distance.}
%\label{fig:moving_distance_small}
%\end{figure}

The probability models $A$, $B$, and $\pi$ are designed to decide the most probable current sidewalk segment. First, the observation probabilities $A$ are computed based on the fact that a GPS location geographically far from the current sidewalk segment is less likely to occur~\cite{newson2009hidden}. An observation probability $P(z_t|r_i)$ for a GPS location $z_t$ thus can be modeled as the probability distribution of the geodetic distance between $z_t$ and $z_{t,i}$. Here $z_{t,i}$ is the geographically closest location from on a sidewalk segment $r_i$ from $z_t$. Let us denote this geodetic distance by $|z_t - z_{t,i}|_{geo}$. Since the geodetic distance represents the GPS positioning error which is known to follow the zero-mean Gaussian{~\cite{chawathe2007segment}}, the observation probability can be written as:

\begin{equation}
\label{eq1}
P(z_t|r_i)=\frac{1}{\sqrt{2\pi}\sigma_z}e^{-0.5(\frac{|z_t-z_{t,i}|_{geo}}{\sigma_z})^2},
\end{equation}

\noindent where $\sigma_z$ is the standard deviation of GPS measurements, which can be obtained empirically. Note that our system regularly updates $\sigma_z$ based on previously measured GPS locations since $\sigma_z$ may change depending on the environment. Now considering a set of GPS locations $Z_t=\{z_1,...z_{\omega}\}$ stored in a sliding window, the observation probabilities $A = P(Z_t|r_i)$ can be defined as follows.

\begin{equation}
\label{eq2}
P(Z_t|r_i) = \frac{\sum_{t=1}^{\omega}P(z_t|r_i)}{\omega},
\end{equation}

\noindent which represents the likelihood that the user is on sidewalk segment $r_i$ given the set of preceding GPS positions $Z_t$.

Next we model the transition probabilities $B$ which define the likelihood that the user transitions to another sidewalk segment. For this, let us define the moving distance between two GPS locations $z_t$ and $z_{t+1}$, denoted by $|z_{t+1}-z_t|_{mov}$. The moving distance refers to the geographic distance between the two GPS locations along the shortest sidewalk trajectory. Fig.~\ref{fig:location_experiment} illustrates the geodetic and moving distance between two GPS locations $z_t$ and $z_{t+1}$. Newson and Krumm noted that the transition probability depends on the difference between the moving distance and the geodetic distance~\cite{newson2009hidden}. More specifically, the transition probability becomes higher when the difference is larger, and vice versa. It was also shown by~\cite{newson2009hidden} that the difference follows the exponential distribution. A trick that we make to account for the slow moving speed of the pedestrian is to use the GPS location measured $\epsilon$ time ago (\emph{i.e.,} $z_{t-\epsilon}$) rather than using the preceding GPS location (\emph{i.e.,} $z_{t-1}$) in calculating the moving and geodetic difference. Now by denoting the distance difference as $\delta = | |z_{t}-z_{t-\epsilon}|_{mov} - |z_{t} - z_{t-\epsilon}|_{geo} |$, we obtain $p(\delta)=\frac{1}{\beta}e^{\frac{-\delta}{\beta}}$.

%\begin{equation}
%p(\delta)=\frac{1}{\beta}e^{\frac{-\delta}{\beta}}.
%\end{equation}

Finally, since the transition probabilities $B$ depend on the distance difference, we obtain: $B = P(r_j|r_i) \approx p(\delta)$. We also note that using Eqs.~\ref{eq1} and~\ref{eq2}, the initial state probabilities $\pi = P(Z_1|r_i)$ can be easily calculated. Given these probability models $A$, $B$, and $\pi$, the proposed map matching algorithm identifies the current sidewalk segment. Once the current sidewalk segment is identified, the ``valid'' region is calculated. Any GPS location that is outside this region is either rejected, or projected onto the region. More precisely, the width of the region is defined as `$\alpha \cdot$ (max walking speed) $\cdot$ (GPS measurement interval)' and the height of it is defined as `$\alpha \cdot$ (the sidewalk width)'. Here, the parameter $\alpha$ is adopted to allow the user to adjust the tolerance to location errors. In our experiments, we used 15m as the threshold to reject a GPS location. If the distance between a measured GPS location and the valid region is greater than 15m, we rejected the GPS location, and if not, the GPS location is projected onto the closest point on the valid region.

\subsection{Improving Energy Efficiency}
\label{sec:energy_engine}

\Figure[t](topskip=0pt, botskip=0pt, midskip=0pt)[width=.9\columnwidth]{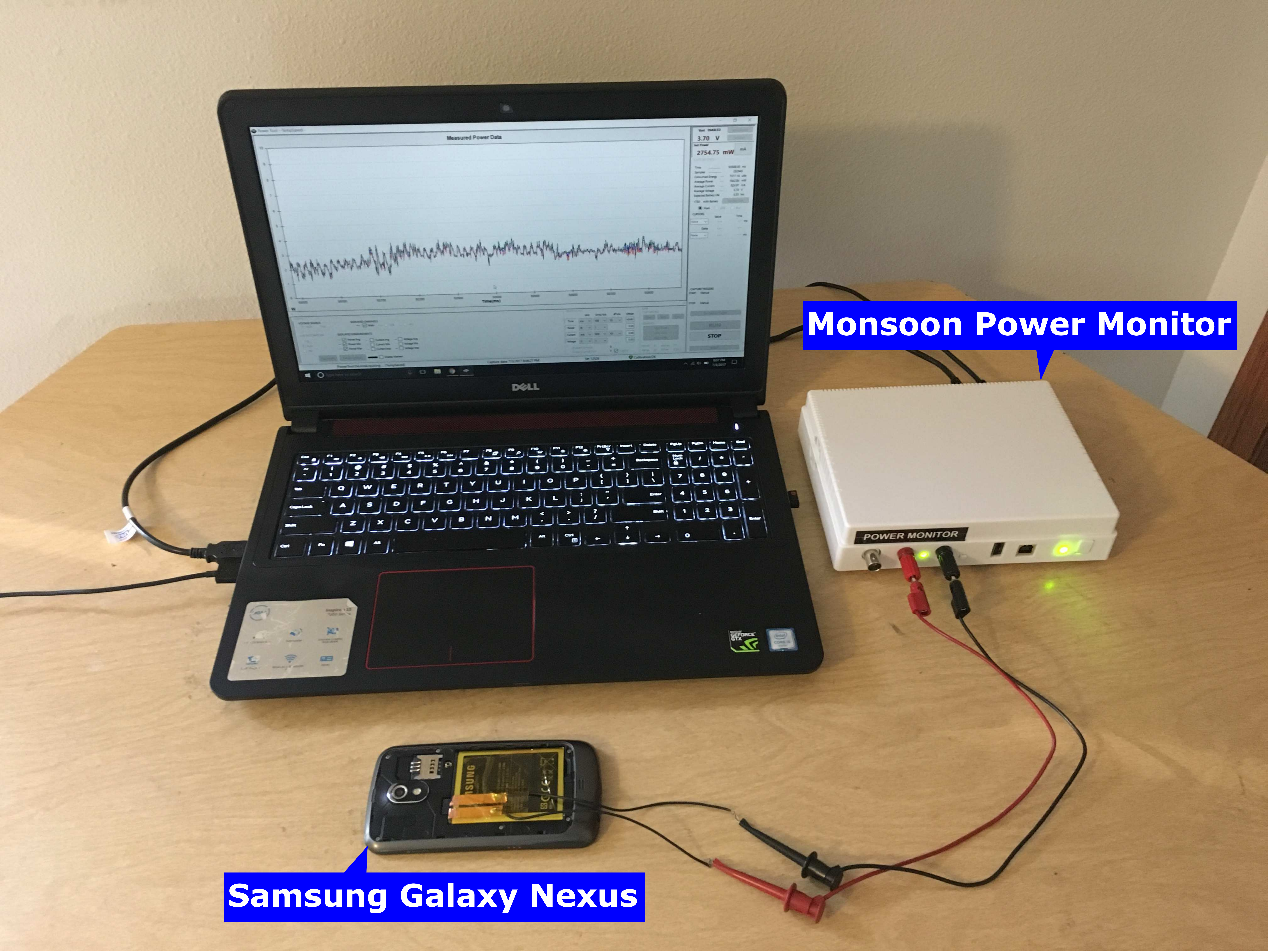}
{Experimental setup for power measurement. The power monitor is directly connected to the phone's battery terminals. \label{fig:gps_power_snapshot}}

\Figure[t](topskip=0pt, botskip=0pt, midskip=0pt)[width=.9\columnwidth]{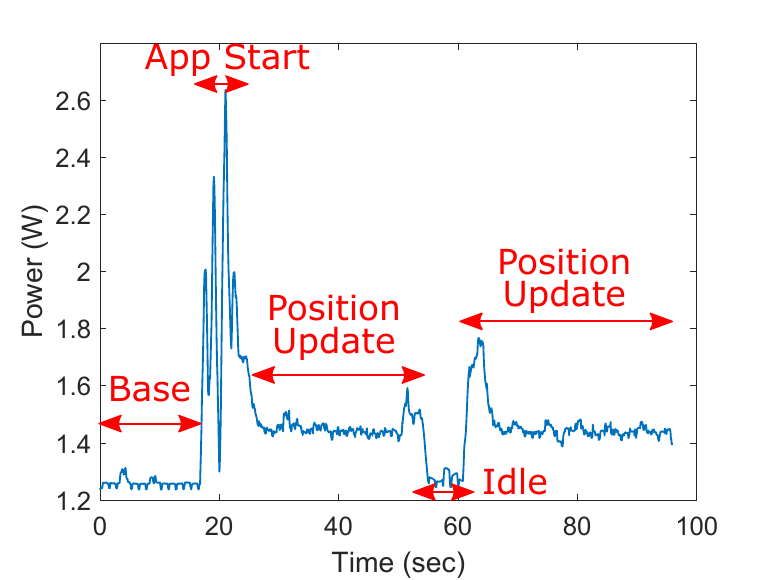}
{Power consumption of the GPS module. The result indicates that significant energy savings are possible by adaptively activating the GPS module. \label{fig:gps_power}}

%\begin{figure}[!htbp]
%\centering
%\includegraphics[width=0.9\columnwidth]{gps_power_snapshot}
%\caption {Experimental setup for measuring power consumption of %$\mathsf{SafePedCross}$.}
%\label{fig:gps_power_snapshot}
%\end{figure}

The GPS module of smartphone is one of the most power hungry sensors~\cite{paek2010energy}. We develop the \textsf{Energy} module to improve the energy efficiency of $\mathsf{SaferCross}$ that heavily utilizes the GPS module. To characterize the energy consumption of the \textsf{Location} module of $\mathsf{SaferCross}$, an experiment was performed using the Monsoon Power Monitor~\cite{monsoon}. Fig.~\ref{fig:gps_power_snapshot} shows the experimental setup. We connected the power monitor's probes to the phone's battery terminals so that the monitor provides current to the phone. And then, a laptop was connected to the power monitor via USB and measured the samples of current drawn and the voltage at a rate of 5KHz.

%\begin{figure}[!htbp]
%\centering
%\includegraphics[width=0.7\columnwidth]{gps_power}
%\caption {Power consumption of $\mathsf{SafePedCross}$.}
%\label{fig:gps_power}
%\end{figure}

Fig.~\ref{fig:gps_power} shows that a large amount of power was consumed for a short period of time when the app was started to load and display the app on the screen. After that, the app used about 1.5 watt for updating  and calibrating the position. An interesting observation was that the GPS module consumed very a small amount of energy, as low as the baseline energy consumption, when the GPS module was put into the sleep mode, indicating that significant energy savings can be achieved by putting the GPS module into the sleep mode. It is worth to note that the sleep mode depends on mobile operating systems. For example, by the sleep mode in Android, we mean that we stop receiving position update, while the GPS module maintains the lock on the acquired satellites, so that when we resume position update, the GPS module does not need to re-acquire and lock on satellites.

\Figure[t](topskip=0pt, botskip=0pt, midskip=0pt)[width=.9\columnwidth]{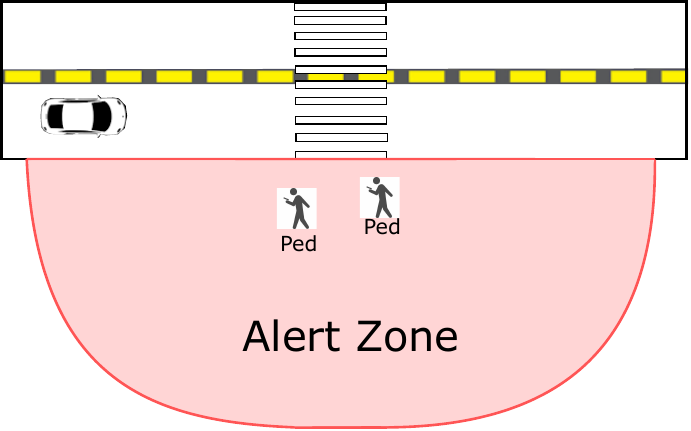}
{An illustration of an alert zone which is a 2D region, the boundary of which is equidistant from a nearest crosswalk. The system components of $\mathsf{SaferCross}$ are activated when the user is within an alert zone.\label{fig:alert_zone}}

%\begin{figure}[!htbp]
%\centering
%\includegraphics[width=.4\columnwidth]{alert_zone}
%\caption {Illustration of alert zone.}
%\label{fig:alert_zone}
%\end{figure}

The \textsf{Energy} module determines dynamically when to turn on the GPS module and when to put it into the sleep mode. To explain the mechanism, we need to define the \emph{alert zone}. The alert zone is a 2D region, the boundary of which is equidistant from a nearest crosswalk (Fig.~\ref{fig:alert_zone}). This alert zone is important for $\mathsf{SaferCross}$ as system components are activated only when the user is within an alert zone. An interesting aspect of this alert zone is that it can also be set up for non-crosswalk areas to prevent accidents for jaywalkers. The basic mechanism for saving energy is to estimate the time when the user will be at an alert zone and put the GPS module into the sleep mode until that time. More specifically, the estimated time is calculated as $\frac{d}{v_{max}}$ where $d$ is the shortest geodetic distance between the current user location and the closest alert zone, and $v_{max}$ is the maximum brisk human walking speed~\cite{walking}. An interesting aspect of the \textsf{Energy} module is that GPS is adaptively controlled in coordination with the provided map and the associated alert zones, considering the user walking direction. The walking direction can be monitored even if the GPS is off based on a known technology~{\cite{roy2014smartphone}} so that the GPS will be turned back on when the user direction is reversed to re-estimate the time.

%Note that the \textsf{Energy} module assumes that the user walks at the fastest brisk walking speed to ensure maximum safety. The reader may notice the tradeoff between the energy efficiency and safety that $\mathsf{SafePedCross}$ offers. More precisely, $\mathsf{SafePedCross}$ can find the exact time when the user is within the alert zone if the GPS is on continuously at the cost of reduced energy efficiency. On the other hand, $\mathsf{SafePedCross}$ may be set to save energy by finding the estimated time of the user being in the alert zone, which may affect the safety if the GPS was turned on late, \emph{e.g.,} if the user was running. $\mathsf{SafePedCross}$ is designed such that the user is allowed to explore and choose an option that is most appropriate to him/her. Another nice aspect of $\mathsf{SafePedCross}$ is that the average walking speed of the user can be used for estimating the arrival time rather than using the maximum brisk walking speed.

\subsection{Detecting Pedestrian Phone Use}
\label{sec:context}

The \textsf{Context} module is developed to effectively detect the user- phone-viewing event. However, detecting the phone-viewing event is hard because it is associated with limited user interactions such as tapping on the phone. There is an approach that utilizes the camera of the phone to detect the phone viewing event by recognizing the user's face/eyes~\cite{dickie2005eyelook}. A limitation of this approach is the privacy concerns. Furthermore, the phone orientation information is insufficient to determine whether the user is viewing the phone or not.

In order to develop a novel approach for detecting the phone-viewing event, we hinge on the observation that when the user views their phone while walking, they tend to try to minimize phone shaking to better read email/text messages, and watch videos. Based on this motivational observation, we quantify phone shaking using the variance of the acceleration magnitude of phone. We then use the quantified data to detect the phone-viewing event. More precisely, given an accelerometer reading $(a_x, a_y, a_z)$ of phone in $x$, $y$, and $z$ directions respectively, we remove random noise using the standard low-pass filtering. As a result, we obtain filtered accelerometer data denoted by $\hat{a_x}, \hat{a_y}, \hat{a_z}$. The magnitude of the acceleration vector $m$ is then calculated as $m=\sqrt{\hat{a_x}^2+\hat{a_y}^2+\hat{a_z}^2}$.

%\begin{equation}
%m=\sqrt{\hat{a_x}^2+\hat{a_y}^2+\hat{a_z}^2}.
%\end{equation}

\Figure[t](topskip=0pt, botskip=0pt, midskip=0pt)[width=.9\columnwidth]{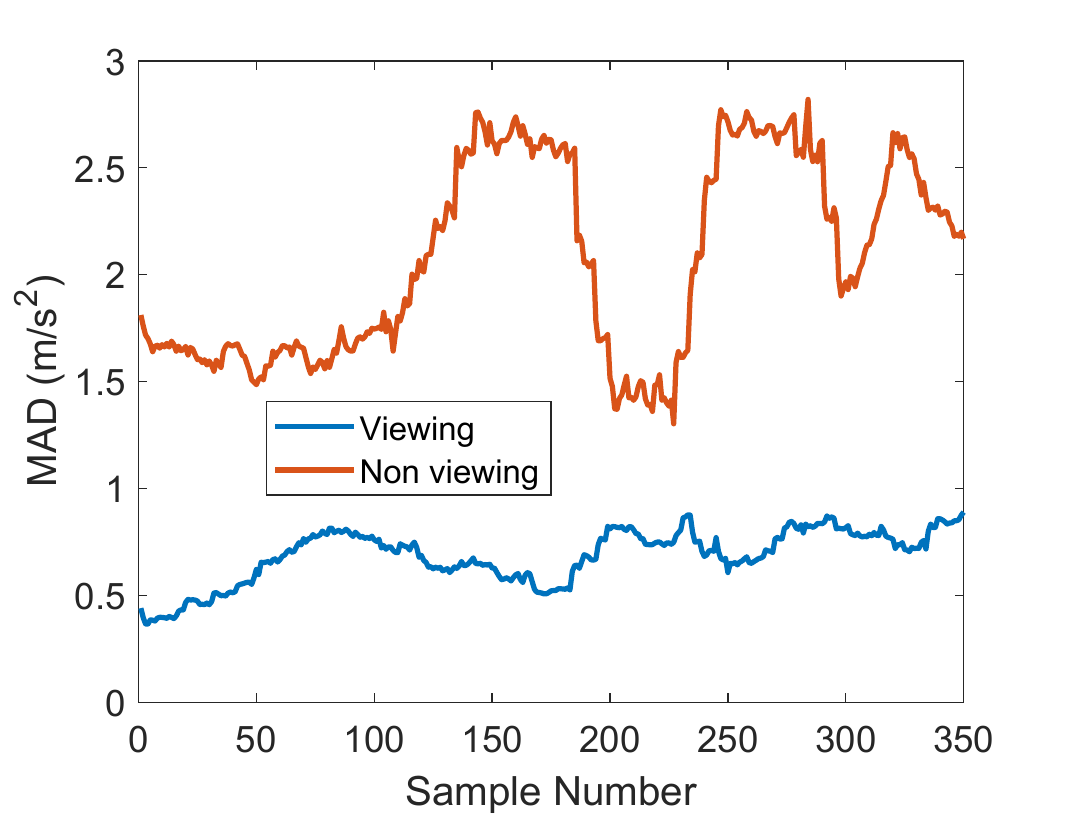}
{MAD values measured with a 10sec window. The two events are clearly distinguished. \label{fig:sliding_window_10_sec}}

\Figure[t](topskip=0pt, botskip=0pt, midskip=0pt)[width=.9\columnwidth]{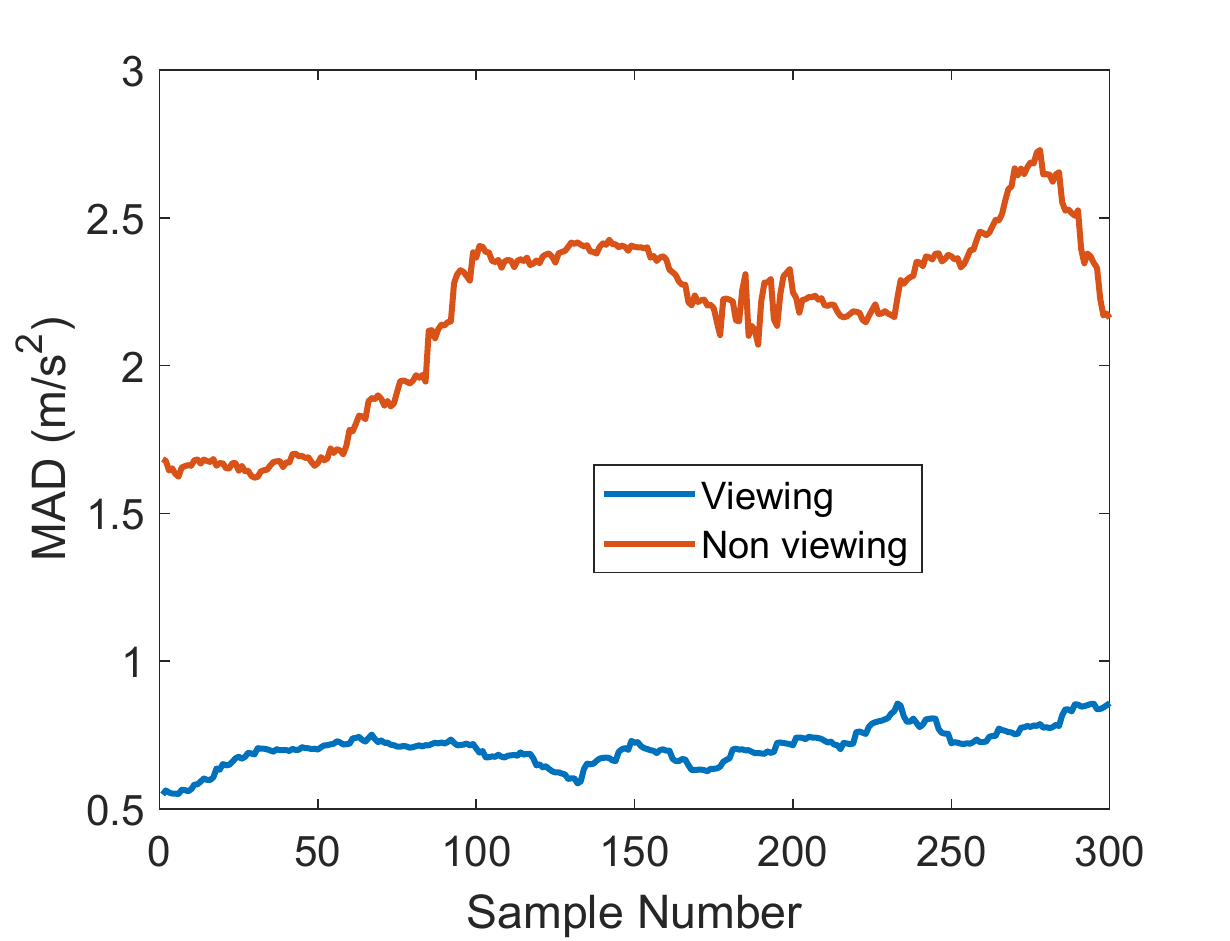}
{MAD values measured with a 20sec window. The two events are more clearly distinguished with a larger window size. \label{fig:sliding_window_20_sec}}

A sliding window $W=\{m_1, m_2,...m_{\phi}\}$ is used to store a sequence of acceleration magnitude values collected over a period of time. The variation of the magnitude values in a window is represented as the mean absolute deviation (MAD) which is used to quantify the shaking of phone. Fig.{~\ref{fig:sliding_window_10_sec}} displays an example of MAD values for both phone-viewing and non-viewing events with a 10-sec sliding window. Leveraging the clear difference between the MAD values of the two events, we design a simple threshold-based method to detect the phone-viewing event. More specifically, a threshold $\Gamma$ is defined as the average of the mid points of MAD values for phone-viewing event and non-phone viewing event. Given training data, \emph{i.e.,} the MAD values for the phone viewing event $X=\{x_1, x_2, ..., x_n\}$, and the MAD values for the non-phone-viewing event $Y=\{y_1, y_2, ..., y_n\}$, the threshold is calculated as $\frac{\sum_{i=1..n} (\frac{y_i + x_i}{2})}{n}$. A more advanced AI-based and dynamic mechanism to determine the threshold is left as a future work.

%\begin{equation}
%\frac{1}{\phi}\sum^{\phi}_{i=1}|m_i - \mbox{mean}(W)|.
%\end{equation}

An experiment was performed to evaluate the feasibility of the proposed approach. Five volunteers participated in this experiment. They were asked to walk with viewing their phones. They were also asked to walk without viewing their phones. Figs.~\ref{fig:sliding_window_10_sec} and~\ref{fig:sliding_window_20_sec} show the results for different sizes of the sliding windows, \emph{i.e.,} 10sec and 20sec, respectively. As it can be seen, MAD values for the phone-viewing scenario were significantly smaller than that for the non-phone-viewing scenario, allowing us to clearly differentiate the two scenarios. The proposed method turns out to be quite accurate with detection accuracy over 90\%.

It is also worth to note that since the \textsf{Context} module is only activated when the user is within an alert zone, and the accelerometer consumes significantly less amount of power than the GPS module, the energy efficiency issue for the \textsf{Context} module is less critical than the \textsf{Location} module.
\subsection{Determining When to Alert the User}
\label{sec:alert_engine}

%The \textsf{Alert} module of $\mathsf{SafePedCross}$ is designed to calculate the collision probability and alerts the user. It calculates the probability of collision utilizing the probability model built upon real world data. Based on the resulting probability of collision, the module sends an alert message to the user and/or the approaching cars.

\Figure[t](topskip=0pt, botskip=0pt, midskip=0pt)[width=.9\columnwidth]{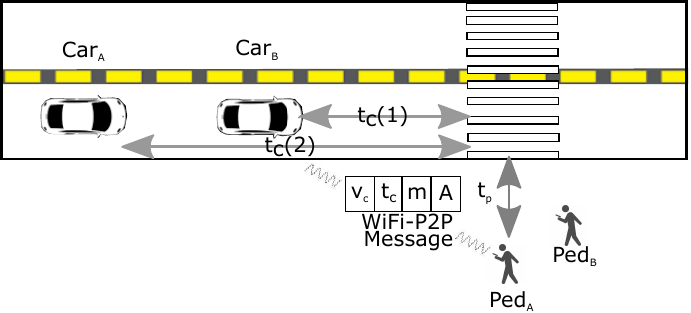}
{An illustration of how the \textsf{Alert} module works. The message exchanged between the pedestrian and the driver contains information to estimate the probability of collision. \label{fig:elert_engine_overview}}

The \textsf{Alert} module of $\mathsf{SaferCross}$ is developed to determine when to alert the user by estimating the collision probability. This module is activated when the user is within an alert zone. It sends a REQ message to approaching vehicles via WiFi Direct. In response to the REQ message, vehicles send a REP message to the user. The REP message contains information required to estimate the collision probability including the vehicle speed $v_c$, vehicle mass $m$, cross-sectional area of the vehicle $A$, and time for the vehicle to reach the crossing denoted by $t_c$ (Fig.~\ref{fig:elert_engine_overview}). The vehicle and the pedestrian keep exchanging these messages to update the collision probability in real time to account for the changing motion of the vehicle and the user.

In estimating the collision probability, the time for the pedestrian to reach at the crossing is calculated, \emph{i.e.,} $t_p = \frac{d_p}{v_p}$ where $d_p$ is the shortest geodetic distance between the pedestrian and the crossing, and $v_p$ is the user walking speed. The Android context API is used to determine $v_p$. Specifically, we use a brisk walking speed{~\cite{walking}} when the API detects that the user is walking; if the user is running, $v_p$ is set to a predetermined running speed. In particular, if the user is not walking or running, the collision probability is not calculated.

It is important to note that to provide the near real-time computation of the collision probability, when the user in an alert zone, the calculation of the current sidewalk segment of the \textsf{Location} module is suppressed, which is based on the observation that the user stays in the same sidewalk segment when she is in the same alert zone.

Given $v_c$, $m$, $t_c$, $A$, and $t_p$, the \textsf{Alert} module is ready to estimate the collision probability. Let $v_c(i)$, $m(i)$, $t_c(i)$, and $A(i)$ be the vehicle speed, vehicle mass, amount of time to reach at a crossing, and cross-sectional area for vehicle $i$, respectively. If $t_p \gg \max (t_c(i)), \forall i$, \emph{i.e.,} if the user is expected to reach at the crossing long after all approaching vehicles have passed, the user is not alerted. On the other hand, if $t_p < \max (t_c(i)), \exists i$, \emph{i.e.,} there is at least one approaching vehicle around the crossing by the time the user reaches at the crossing, the module estimates the collision probability. Consequently, if the following two conditions are satisfied, an alert message is generated for the user: (a) The pedestrian is walking/running while viewing the phone; (b) The probability of collision is greater than a threshold. Note that the estimation of the collision probability is continually updated as the REQ and REP messages are kept being exchanged between vehicles and the user. Thus, if there is any new vehicle within the range of WiFi Direct, the collision probability for that new vehicle will be calculated and updated.

More details are presented on how the collision probability is calculated. First, we define a term `user warning time' denoted by $t_{warning}  = \min (t_c(i)) - t_p$ that represents the amount of time allowed for the driver to avoid an accident after he sees the pedestrian who is about to cross the street. And then, the collision probability is estimated as $P(t_{delay}+t_{react}+t_{skid} > t_{warning})$, where $t_{delay}$ is the round-trip message delay for a single-hop 802.11 link. $t_{react}$ is the driver reaction delay, and $t_{skid}$ is the amount of time from the point when the driver applies brakes until the car completely stops. If the sum of these time delays is greater than $t_{warning}$, the likelihood of collision is deemed high. In particular, we disregard the WiFi Direct connection establishment time since the connection has been already established before the first alert message is sent from the user to the approaching cars.

%\begin{equation}
%P(t_{delay}+t_{react}+t_{skid} > t_{warning}),
%\end{equation}

More specifically, $t_{delay}$ is empirically obtained as the pedestrian continuously exchanges messages with approaching vehicles, \emph{i.e.,} $t_{delay}$ is the average of measured round-trip message delays. In calculating $t_{react}$, we leverage the observation that the log-normal probability model fits the driver reaction time well~\cite{taoka1989brake}. Thus, $t_{react}$ is defined based on the log-normal distribution{~\cite{feller2008introduction}} as follows:

\begin{equation}
\label{eq:react}
f(x|\mu,\sigma)=\frac{1}{x\sigma\sqrt{2\pi}}e^{\frac{-(\ln x-\mu)^2}{2\sigma^2}},
\end{equation}

\noindent where we select the mean and standard deviation of the driver reaction time as $\mu=1.14$ and $\sigma=0.32$, respectively according to the experimental data collected by Gaziz \emph{et al.}{~\cite{gazis1960problem}}. To calculate $t_{skid}$, we first compute $d_{skid}$ that is the distance that a car moved until it is completely stopped after brakes are applied as follows: $d_{skid}=\frac{mv_p^2}{2f}$, where $m$ is the vehicle mass, and $v_p$ is the vehicle speed, which we obtain from the REP message. $f$ is the resistance force, which is calculated based on the model proposed by Ho \emph{et al.}{~\cite{ho2016wisafe}}.

%\begin{equation}
%d_{skid}=\frac{mv_p^2}{2f},
%\end{equation}

\begin{equation}
f=\mu_kmg + \frac{\rho AC_d v_r^2}{2} + f_0,
\end{equation}

\noindent where $\rho$ is the density of air, $A$ is the cross-sectional area of the vehicle, $C_d$ is the drag coefficient, $v_r$ is the speed of the vehicle relative to the air, and $f_0$ is the other resistance force. In our experiments performed on a sunny day on a good conditioned road with Volkswagen Passat 2013, we used the parameter: $m=1400kg$, $\mu_k=0.8$, $A=2.7m^2$, $C_d=0.25$, $\rho=1.23kg/m^3$ according to \cite{ankrum1992ivhs}\cite{rivers2006evidence}\cite{hugemann2002driver}. $v_r$ was approximated as the current vehicle speed $v$ due to the slow wind speed. Thus, $t_{skid}=\frac{d_{skid}}{v}$.

Once $t_{delay}$, $t_{skid}$, and $t_{warning}$ are known, the collision probability can be written as: $P(t_{react} > t_{warning} - t_{delay} - t_{skid})$ which can be calculated leveraging the fact that $t_{react}$ follows the log-normal distribution specified in Eq.~\ref{eq:react}. Note that we are very careful in sending an alert message to approaching cars. The reason is that alert messages may disturb safe driving. In designing the \textsf{Alert} module, thus, we give an emphasis on alerting the user first in an hope that the user will stop and look up when they receive the alert message. However, if the user ignores the alert message (\emph{e.g.,} by clicking the cancel button), an alert message is eventually sent to the driver.

\subsection{Enabling Communication Between Pedestrian And Cars}
\label{sec:communication}

To enable direct communication between the user and approaching cars, WiFi Direct is used. WiFi Direct is a standard designed by the WiFi alliance to facilitate device-to-device (D2D) communication between nearby devices without involving an access point~\cite{camps2013device}. In WiFi Direct, devices communicate by establishing a group. One of them is the group owner (GO), and the others are the group members (GM). These roles are negotiated by the devices in the device discovery phase. The GO implements the AP-like functionality, and the GMs act like clients. Specifically, the GO advertises to its GMs and allows new GMs to join the group. The GO runs a Dynamic Host Configuration Protocol (DHCP) server to provide IP addresses to joining GMs after going through the WiFi Protected Setup (WPS) phase.

In $\mathsf{SaferCross}$, the pedestrian is the GO, and approaching cars are the GMs. An approaching car scans a predefined channel to search for the GO. Once the GO is discovered, the car joins the group immediately.  There are two main challenges. The first one is that these scanning and negotiation processes take too much time. The literature shows that it can take about 8 to 9 seconds~\cite{camps2013device}. Fortunately, WiFi Direct provides the \emph{autonomous mode} in which the negotiation process is not required as the GO is predetermined. The autonomous mode fits perfectly with $\mathsf{SaferCross}$ because there are clear roles, \emph{i.e.,} the user and the cars. Our experiments show that the average time to form a group in the autonomous mode is 2.8 seconds, which coincide with the results of the previous research~\cite{camps2013device}.

Another challenge is that WiFi Direct is essentially designed to support 1-to-1 or 1-to-many communication. For example, consider Fig.~\ref{fig:elert_engine_overview} in which the user $Ped_A$ forms a group (1-to-many) with two cars $Car_A$ and $Car_B$. However, there may be other users around, say $Ped_B$ in this figure who wants to communicate with the cars. Basically, the challenge is how to allow the GMs (cars) to join more than one groups. According to the WiFi Direct Specification, operating GMs for more than one groups is not precluded, but the implementation is not described~\cite{wifi-direct-spec}.

We address this challenge by allowing the user to overhear on the operating channel for a very brief moment, if the user is not the GO. For example, $Ped_B$ overhears the message exchanges between $Ped_A$ and the cars. $Ped_B$ then finds that the cars have already formed a group with $Ped_A$. Since there is already a group, $Ped_B$ joins as a GM and communicates with $Ped_A$ instead of the cars. Now the trick is that the REP messages received from the GO (\emph{i.e.,} $Ped_A$) are forwarded to $Ped_B$. As a result, although $Ped_B$ is not the GO, it still can receive the vehicle information that it needs to compute the collision probability. Essentially, the proposed solution effectively establishes the virtual n-to-n communication based on a single group.

\section{Experimental Results}
\label{sec:experimental_results}

%\subsection{Experimental Setup}
%\label{sec:experimental_setup}

%\begin{figure}[!htbp]
%\centering
%\includegraphics[width=.9\columnwidth]{feasibility_pdr}
%\caption {Packet delivery rates for varying distance for WiFi Direct.}
%\label{fig:feasibility_pdr}
%\end{figure}

\Figure[t](topskip=0pt, botskip=0pt, midskip=0pt)[width=.9\columnwidth]{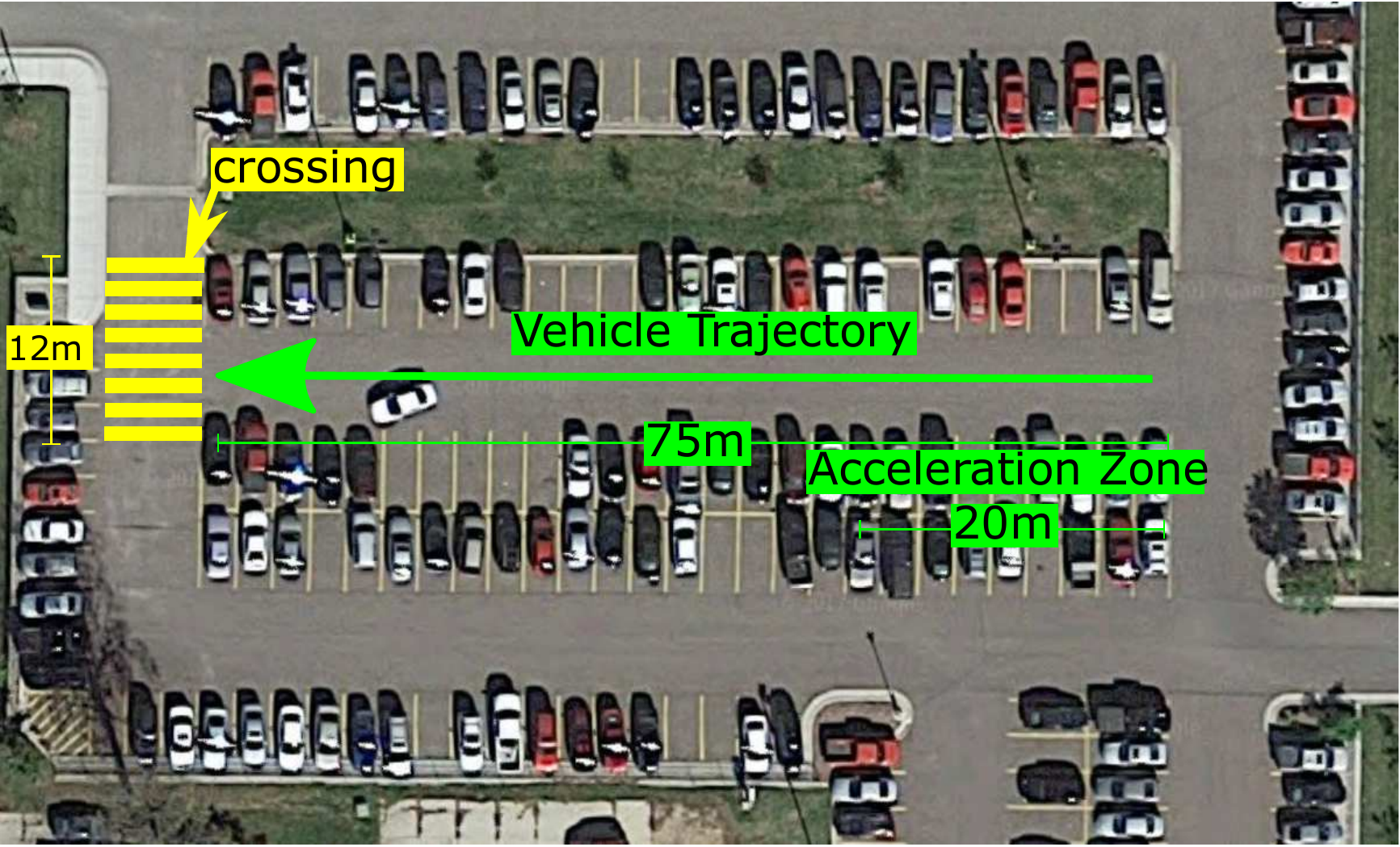}
{A department parking lot used as an experimental site. The acceleration zone is used to reach the desired vehicle speed. \label{fig:test_site}}

\Figure[t](topskip=0pt, botskip=0pt, midskip=0pt)[width=.9\columnwidth]{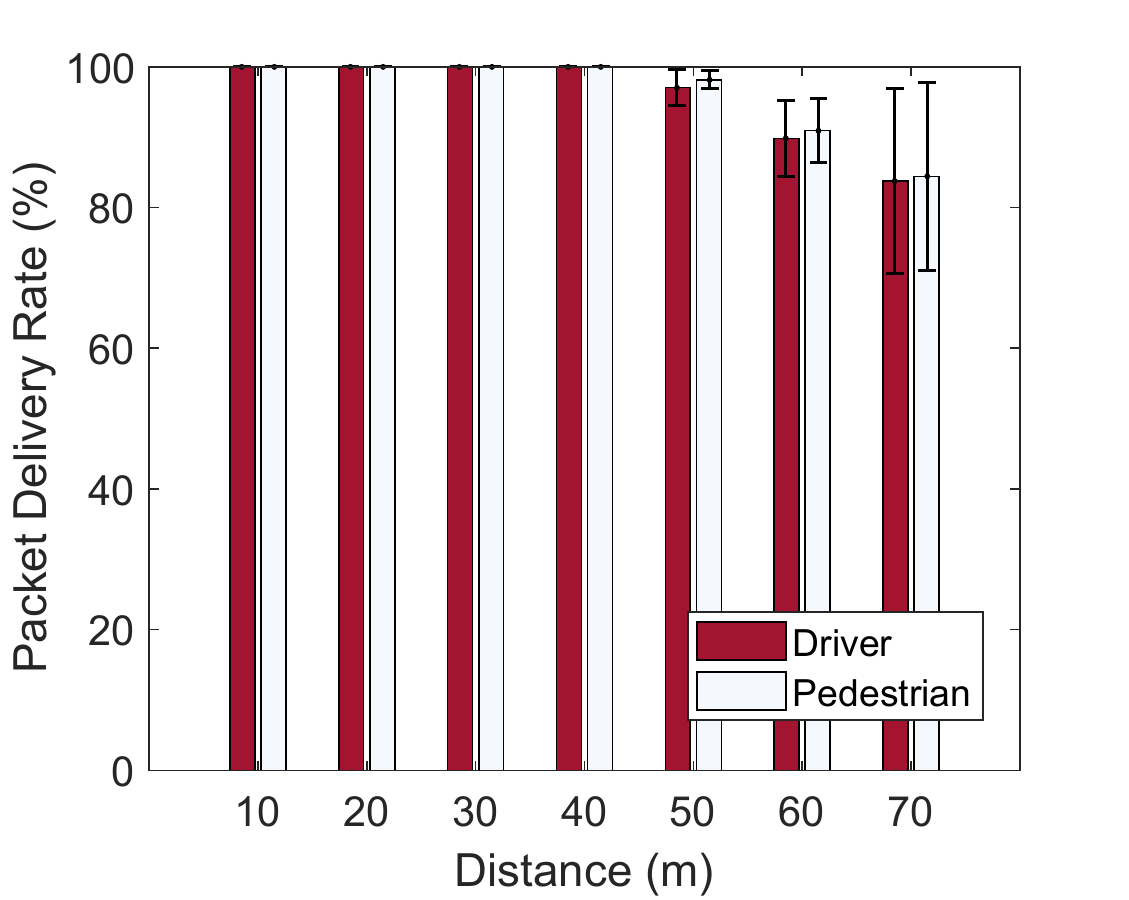}
{PDR for WiFi Direct in the experimental site. The length of the road segment was determined based on the measured PDR. \label{fig:feasibility_pdr}}

\Figure[t](topskip=0pt, botskip=0pt, midskip=0pt)[width=.9\columnwidth]{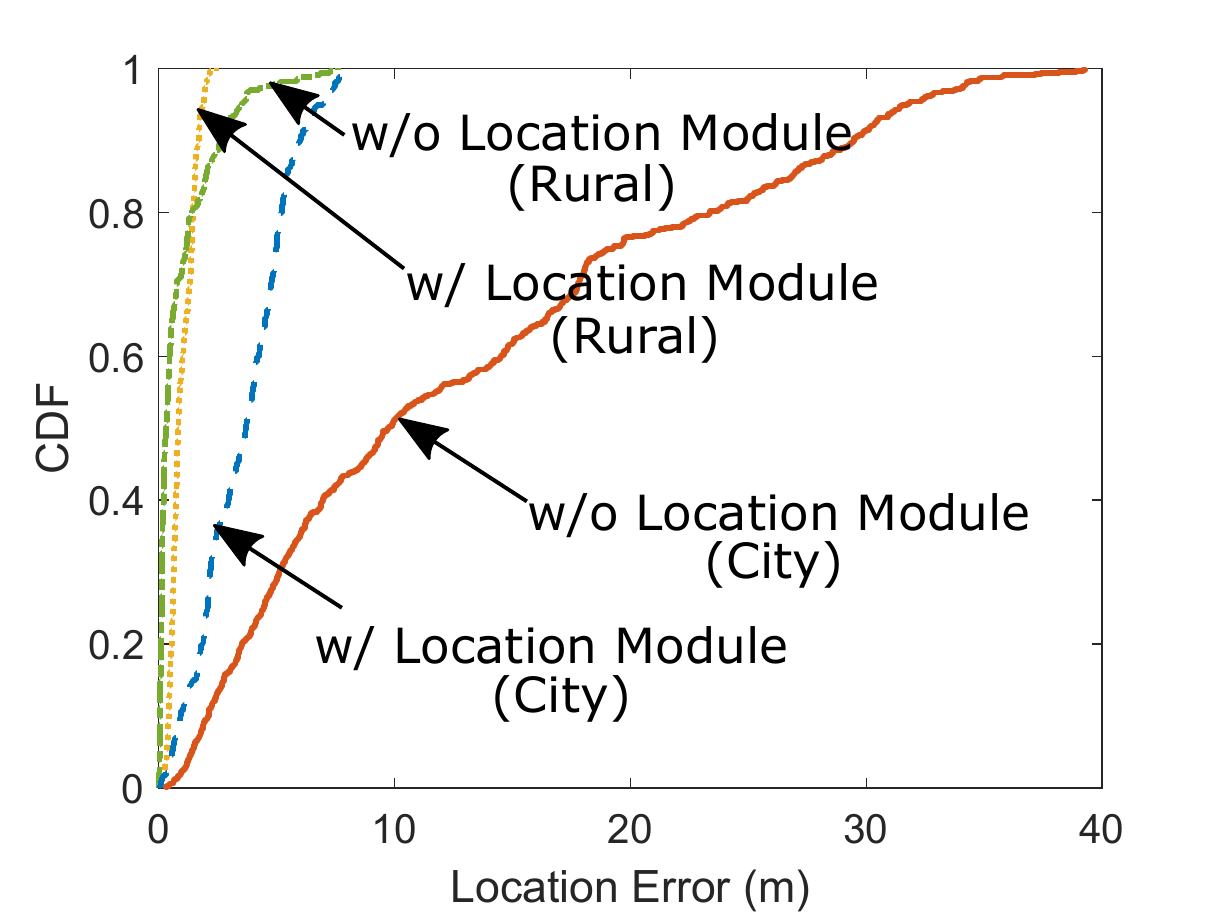}
{CDF of measured location error. The results show that the \textsf{Location} module significantly improves the positioning accuracy. \label{fig:cdf_of_gps_err}}

We implemented $\mathsf{SaferCross}$ on a Samsung Galaxy S6 which is equipped with 1.5GHz octa-core processor and 3GB RAM running on Android 5.0. We performed experiments in a department parking lot (Fig.~\ref{fig:test_site}). To characterize the experimental environment, the packet delivery rates (PDR) were measured for both Pedestrian $\rightarrow$ Driver, and Driver $\rightarrow$ Pedestrian. The average PDR was over 90\% when the distance between the car and the pedestrian was smaller than 60m (Fig.~\ref{fig:feasibility_pdr}). Based on the results, the length of the emulated road segment was set to 75m including the 20m acceleration zone.

%The length of the road segment is long enough for avoiding an accident -- the literature shows that if the distance between a crossing and an approaching car is greater than 40m, the driver can avoid an accident if the vehicle speed is 60km/h or less (37mph)~\cite{ho2016wisafe}.

A driver ran $\mathsf{SaferCross}$ in the driver mode. A Volkswagon Passat'13 was used which moved along the 75m road segment. The driver was asked to accelerate the car to reach the desired vehicle speed in the 20m acceleration zone (Fig.~\ref{fig:test_site}). After reaching the desired speed, the vehicle's cruise control was used to maintain the same speed. Another participant was asked to act as a pedestrian with $\mathsf{SaferCross}$ in the pedestrian mode and walk toward the crossing. To ensure safety, we made sure that the pedestrian always stops at the crossing.

%\begin{figure}[!htbp]
%\centering
%\includegraphics[width=.9\columnwidth]{cdf_of_gps_err}
%\caption {CDF of location error.}
%\label{fig:cdf_of_gps_err}
%\end{figure}

We performed the module-level test first to evaluate the performance of individual system components. We then conducted the integrated test to evaluate the overall performance of $\mathsf{SaferCross}$. The main parameter used for this experiment was the vehicle speed, and we used the `user warning time' $t_{warning}$ as the main metric because it effectively measures the performance of $\mathsf{SaferCross}$ as a whole. Specifically, we can get the accurate user warning time only if all other system components work correctly and the interplay of these components functions effectively. The measured user warning time was compared with the ground truth data.

The experimental environment serves effectively the purpose of evaluating the performance of $\mathsf{SaferCross}$ in comparison with conducting the experiment in real roads. The individual module test can be done readily without accounting for real traffic conditions. Also, the integrated test would effectively approximate the performance of $\mathsf{SaferCross}$ in real roads, because eventually the pedestrian maintains communication only with the foremost vehicle for estimating the collision probability regardless of the traffic of approaching vehicles, and the vehicle used in the experiment effectively represents the foremost vehicle. It should be noted, however, that in order to understand better the effect of other real-world factors such as obstacles, weather conditions, lighting conditions, and human factors, performing experiments in real roads would be valuable. Due to the space limitation and restricted access to public roads, we had to leave the extension of the experiment as future work.

%In this section, however, we demonstrate that the performance of the system modules has been successfully tested both individually and holistically, validating the potential of the system components for being adopted by future mobile systems for pedestrian safety.

\subsection{Positioning Accuracy}
\label{sec:loc_exp}

%\begin{figure}[!htbp]
%\begin{minipage}[b]{0.48\columnwidth}
%\centering
%\includegraphics[width=\columnwidth]{cdf_of_gps_err}
%\caption {CDF of location error.}
%\label{fig:cdf_of_gps_err}
%\end{minipage}
%\hspace{1mm}
%\begin{minipage}[b]{0.48\columnwidth}
%\centering
%\includegraphics[width=\columnwidth]{cdf_of_loc_engine_err}
%\caption {CDF of location error after calibration.}
%\label{fig:cdf_of_loc_engine_err}
%\end{minipage}
%\end{figure}

Positioning accuracy was measured in both rural and city areas without using the \textsf{Location} module first. Five different trajectories were used in each area. In each experiment, a participant was asked to walk along the trajectories to measure the GPS locations and calculate the location errors. Specifically, the location error was defined as the shortest geodetic distance from the measured GPS location to the ground-truth trajectory. Fig.~\ref{fig:cdf_of_gps_err} shows the cumulative distribution graph of the location errors for both the rural and city environments. The mean location error for the rural area was 0.9m. The location error for the city area was significantly greater than the rural area as 12.9m due to many obstacles that disturbed reception of signals from satellites.

%\begin{equation}
%\min_{0 \le i \le n}\{\mbox{dist}(p, \ell_i)\},
%\end{equation}

%This section explains how much the proposed \textsf{Location} module can improve the localization accuracy in these experimental environments. We also discuss if the resulting localization accuracy is sufficiently good enough to run $\mathsf{SafePedCross}$.

To improve the localization accuracy, the \textsf{Location} module was activated, and the experiment was performed under the same conditions. In particular, other system modules were turned off in order to focus on evaluating the effectiveness of the \textsf{Location} module. The results show that the \textsf{Location} module significantly reduced the location errors (Fig.~\ref{fig:cdf_of_gps_err}). The average location errors for the rural and city areas after applying the \textsf{Location} module were 0.8m and 3.5m, respectively. Although the improvement was not significant for the rural area because the location accuracy was already high without the \textsf{Location} module, the module successfully decreased the location error by 72\% in the urban area. Of course, an average error of 3.5m in the urban area is not negligible; yet, it can be compensated by increasing the size of the alert zone, \emph{i.e.,} by providing an alert message to the user a bit early.

\subsection{Energy Efficiency}
\label{sec:energy_exp}

\Figure[t](topskip=0pt, botskip=0pt, midskip=0pt)[width=.9\columnwidth]{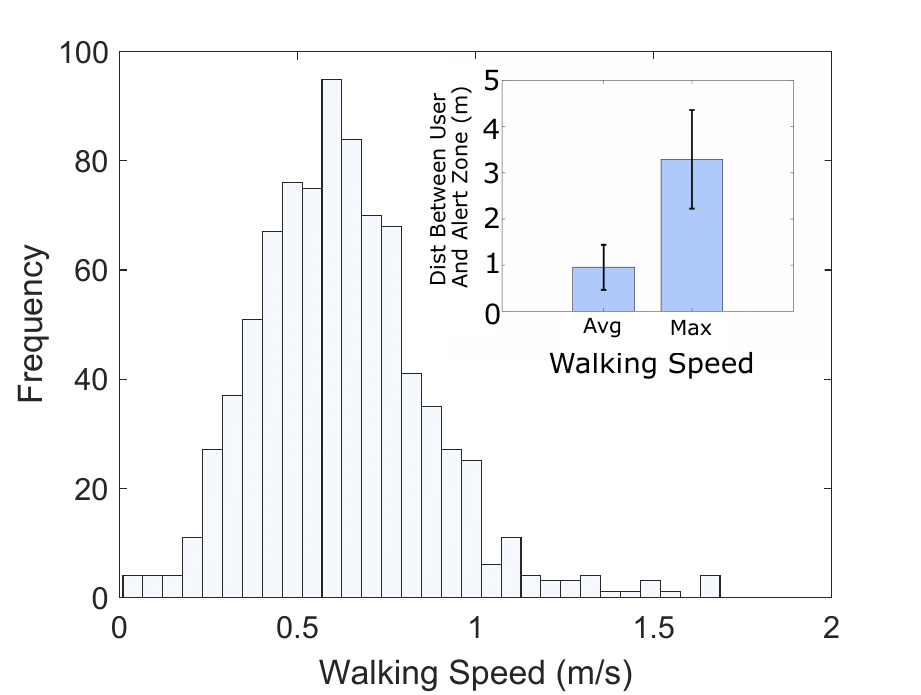}
{The histogram of walking speed data. The data was used by the \textsf{Energy} module in the experiments.\label{fig:walk_histogram}}

We evaluate the performance of the \textsf{Energy} module focusing on two key questions: (1) Is GPS reactivated timely, and (2) how much energy savings are achieved. To answer the first question, we measured the shortest geodetic distance between the pedestrian and the alert zone when GPS was reactivated by the \textsf{Energy} module. If the distance is small, it means that GPS is reactivated timely. This experiment was performed with both the actual walking speed and the brisk walking speed. The actual walking speed was measured by asking the participant to walk for 5mins. Fig.~\ref{fig:walk_histogram} shows the histogram of the actual walking speed. We then calculated the average walking speed and integrated it into the \textsf{Energy} module.

%\begin{figure}[!htbp]
%\centering
%\includegraphics[width=.99\columnwidth]{walk_histogram}
%\caption {Histogram of measured walking speed.}
%\label{fig:walk_histogram}
%\end{figure}

%\begin{figure}[!htbp]
%\begin{minipage}[b]{0.48\columnwidth}
%\centering
%\includegraphics[width=\columnwidth]{walk_histogram}
%\caption {Histogram of measured walking speed.}
%\label{fig:walk_histogram}
%\end{minipage}
%\hspace{1mm}
%\begin{minipage}[b]{0.48\columnwidth}
%\centering
%\includegraphics[width=\columnwidth]{efficiency_of_energy_engine}
%\caption {Performance of \textsf{Energy} module.}
%\label{fig:efficiency_of_energy_engine}
%\end{minipage}
%\end{figure}

%\begin{figure}
%\minipage{0.52\textwidth}
%  \includegraphics[width=\linewidth]{gps_power_measurement}
%\caption {A scenario for energy consumption measurement.}
%\label{fig:gps_power_measurement}
%\endminipage\hfill
%\minipage{0.46\textwidth}
%  \includegraphics[width=\linewidth]{gps_power_result}
%\caption {Energy consumption of \textsf{Location} module.}
%\label{fig:gps_power_result}
%\endminipage
%\end{figure}

\Figure[t](topskip=0pt, botskip=0pt, midskip=0pt)[width=.9\columnwidth]{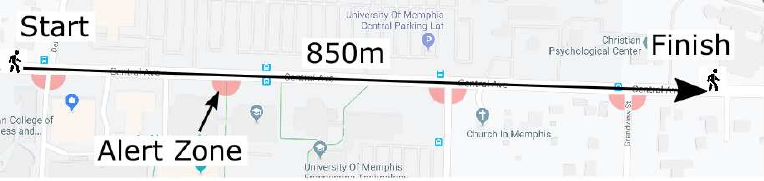}
{A scenario designed for measuring energy savings by using the \textsf{Location} module. The alert zones are set up around the crossings. \label{fig:gps_power_measurement}}

The pedestrian was asked to walk from 30 meters away from the alert zone toward the alert zone. We then measured the geodetic distance when GPS was reactivated. The results are shown in Fig.~\ref{fig:walk_histogram}. The shortest geodetic distance
between the pedestrian and the alert zone was about 1m when the average of the actual walking speed was used by the module. In contrast, when the module used the brisk walking speed, the shortest geodetic distance was about 4m. The results may seem that using the actual walking speed for estimating the time to reactive GPS is better. However, we note that reactivating GPS several seconds early actually would not affect much the energy efficiency, and in fact, it could improve the safety of the pedestrian since other system components are activated several seconds early to allow for more time for the pedestrian to respond to the alert message.

\Figure[t](topskip=0pt, botskip=0pt, midskip=0pt)[width=.9\columnwidth]{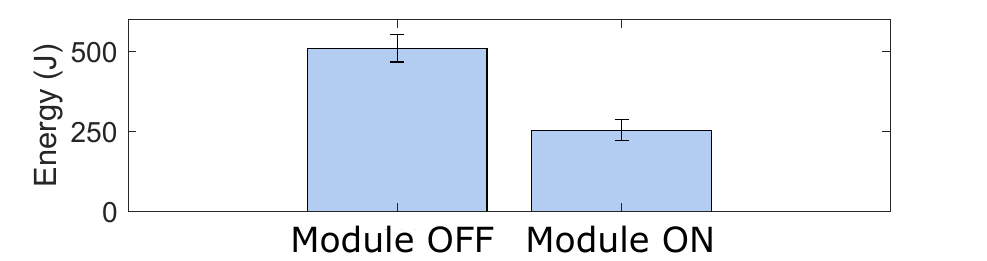}
{Energy savings by using the \textsf{Energy} module. The results indicate that significant energy savings are achieved by using the \textsf{Energy} module.\label{fig:gps_power_result}}

\Figure[t](topskip=0pt, botskip=0pt, midskip=0pt)[width=.9\columnwidth]{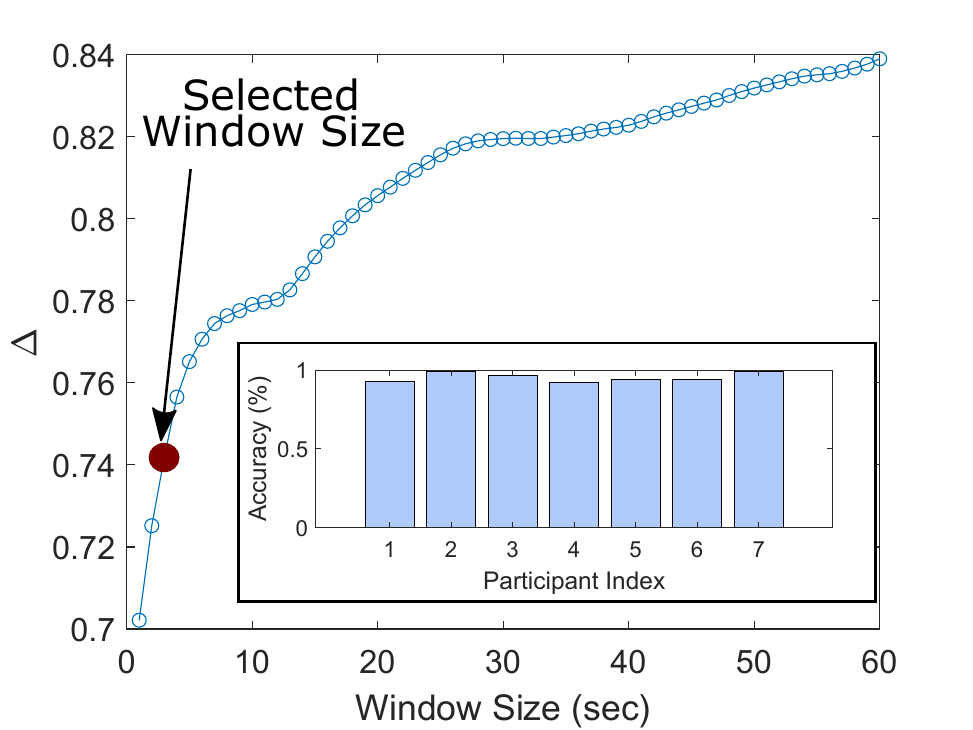}
{Effect of the window size on the accuracy of detecting the phone viewing event. The accuracy is over 90\% for all participants. \label{fig:context_window_size}}

Assuming the brisk walking speed, we evaluated energy savings resulting from the \textsf{Energy} module.

We then performed experiments to understand how much energy savings can be achieved. In this experiment, we used the brisk walking speed. Specifically, we created 8 alert zones along a 850m sidewalk and asked the pedestrian to walk along the sidewalk repeating 5 times (Fig.~\ref{fig:gps_power_measurement}). In this experiment, only the \textsf{Location} module was turned on to provide the calibrated location information. We then measured energy consumption with and without the \textsf{Energy} module. The results are depicted in Fig.~\ref{fig:gps_power_result} demonstrating that \textsf{Location} module decreased energy consumption by 50.2\%.

\subsection{Context Detection Accuracy}
\label{sec:context_exp}

We measured the accuracy of detecting the phone viewing event. The accuracy is defined as the sum of true positives and true negatives divided by the total number of event detection. In this experiment, 7 volunteers participated. They were asked to walk with and without viewing their phones for 10mins each to collect the training data.

%\begin{figure}[!htbp]
%\centering
%\includegraphics[width=.9\columnwidth]{context_window_size}
%\caption {Effect of window size on differentiating the phone-viewing and non-phone-viewing events.}
%\label{fig:context_window_size}
%\end{figure}

The size of the sliding window is determined before measuring the accuracy. We should choose the sliding window size that makes clear distinction between the phone viewing and non-phone viewing events. To quantify how well the two events are differentiated, a new metric $\Delta = \frac{\sum_{i=1..n} (|y_i - x_i|)}{n}$ is defined, where $X=\{x_1, x_2, ..., x_n\}$ and $Y=\{y_1, y_2, ..., y_n\}$ are MAD values for the phone-viewing and non-phone-viewing events, respectively. Thus, higher $\Delta$ values are preferred because it will lead to higher event detection accuracy due to the fact that the two events are more clearly differentiated. However, note that higher accuracy does not necessarily mean higher $\Delta$ values because the phone-viewing event will be detected even if the difference between the MAD values of the two events is small, which is the reason why this new metric is defined to determine the window size.

To decide an appropriate window size, we measured $\Delta$ by varying the window size. Fig.~\ref{fig:context_window_size} shows the results which indicate that using a larger window distinguishes the two events better because of more samples contained in the window. A downside of using a large window size is, however, the increased delay to fill up the window with samples. An interesting observation is that even if we use a small window, $\Delta$ does not decrease too much. For example, $\Delta$ for the window size of 3sec is only 12\% smaller than that for the window size of 60sec. In this experiment, we decided to use the window size of 3sec.

%\begin{figure}[!htbp]
%\centering
%\includegraphics[width=.7\columnwidth]{context_threshold}
%\caption {Detection accuracy of \textsf{Context} module.}
%\label{fig:context_threshold}
%\end{figure}

%We also determine the threshold $\Gamma$ for this experiment. Recall that if a measured MAD value is greater than the threshold, the phone viewing event is detected. We calculate the threshold as the average of the mid points of MAD values for phone-viewing event and non-phone viewing event. The collected training data were used to calculate the threshold. Precisely, given $X=\{x_1, x_2, ..., x_n\}$ and $Y=\{y_1, y_2, ..., y_n\}$, the threshold is calculated as $\frac{\sum_{i=1..n} (\frac{y_i + x_i}{2})}{n}$. A more advanced AI-based and dynamic mechanism to determine the threshold is left as an interesting future work.

With the window size, we measured the event detection accuracy. This time the volunteers were asked to walk with and without viewing their phones for another 5 mins. Fig.~\ref{fig:context_window_size} shows the results. The accuracy was varied depending on the individual as each participant had a different walking style. However, it can be noted that the accuracy for all participants was greater than 90\%, validating that the \textsf{Context} module effectively detects the phone viewing event.

\subsection{Integrated Test}

\Figure[t](topskip=0pt, botskip=0pt, midskip=0pt)[width=.9\columnwidth]{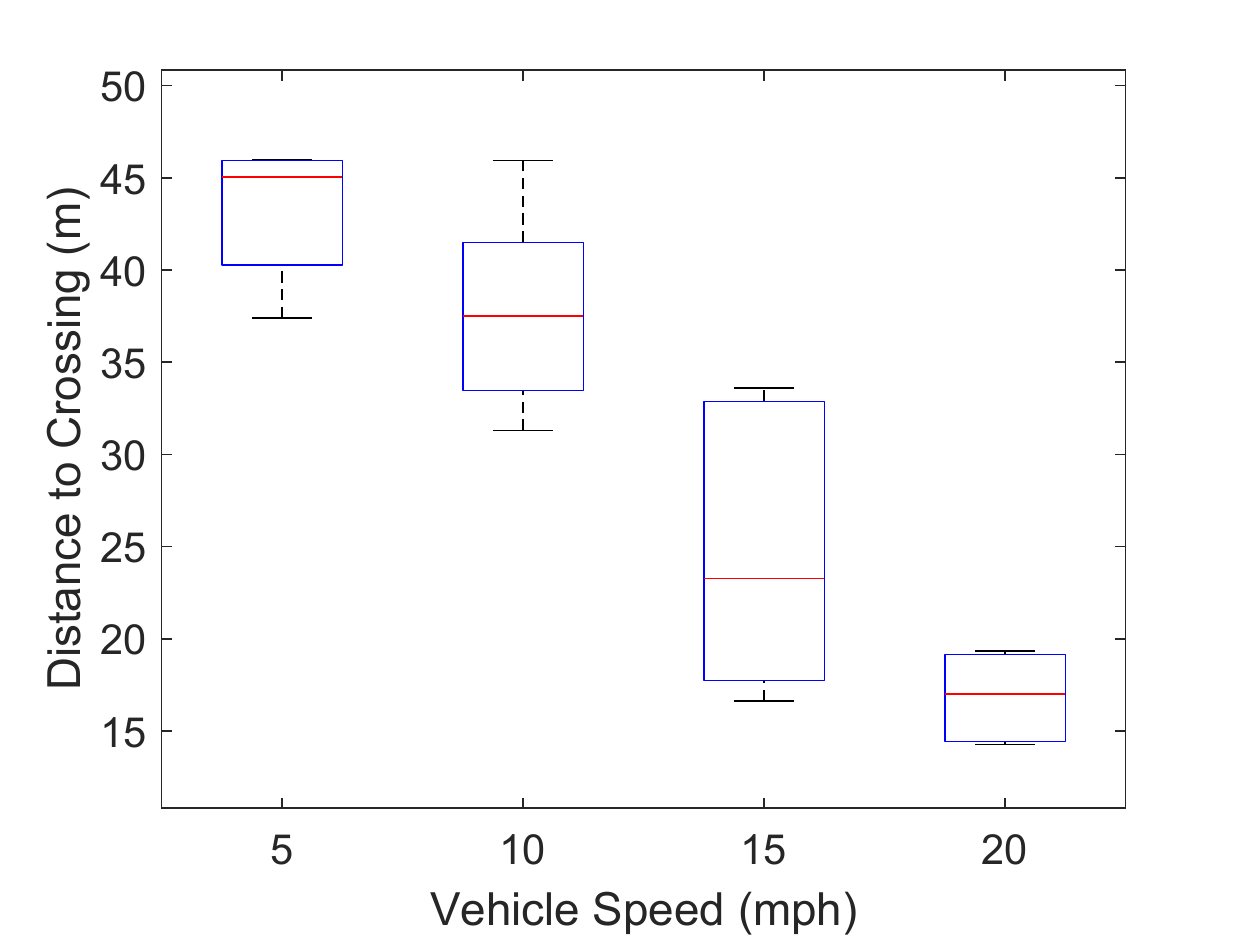}
{The ground-truth distance to the crossing. It varies due to the nondeterministic nature of the user walking speed, GPS locations, and message delay. \label{fig:driver_warning_vehicle_speed}}

\Figure[t](topskip=0pt, botskip=0pt, midskip=0pt)[width=.9\columnwidth]{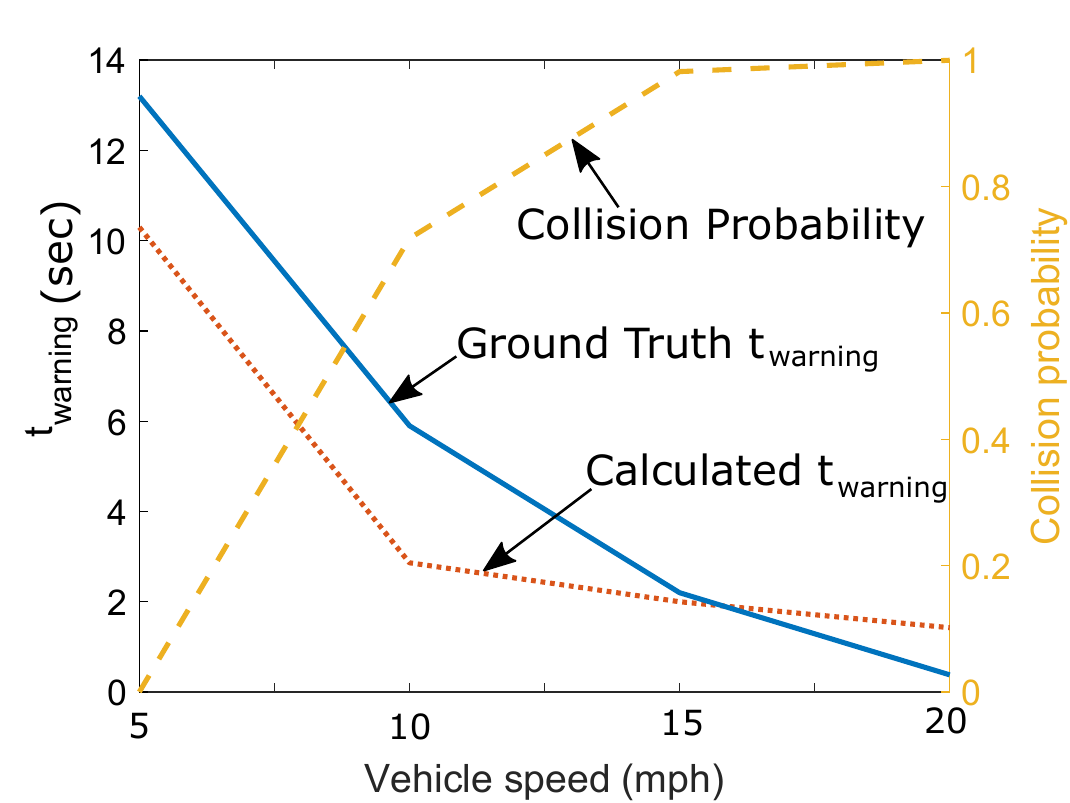}
{User warning time and collision probability. The estimated user warning time is very close to the ground-truth value. The difference can be used to calibrate $\mathsf{SaferCross}$ for better pedestrian safety. \label{fig:user_warning_threshold}}

%\label{sec:alert_exp}
%\begin{figure}[!htbp]
%\centering
%\includegraphics[width=.9\columnwidth]{driver_warning_vehicle_speed}
%\caption {Ground truth distance to the crossing when the vehicle received the alert message.}
%\label{fig:driver_warning_vehicle_speed}
%\end{figure}

We perform an integrated test to evaluate the performance of $\mathsf{SaferCross}$ as a whole by putting together all the individual modules. We use the user warning time $t_{warning}$ as a main metric for performance evaluation based on the observation that accurate $t_{warning}$ can only be obtained if all system modules perform effectively. Specifically, in this experiment, $t_c$ and $t_p$ were recorded to calculate the user warning time, \emph{i.e.,} $|t_c - t_p|$. However, measured $t_{warning}$ may be different from the ground-truth time to collision denoted by $t_{warning}^{GT}$ due to various factors such as the positioning error (for both the car and the pedestrian), processing delay, and transmission delay for delivering the warning message to the driver. We focus on capturing the difference, \emph{i.e.,} $|t_{warning} - t_{warning}^{GT}$ in evaluating the overall system performance.

In this experiment, a participant was asked to walk toward the crossing while viewing his smartphone. At the same time, a driver was asked to drive a car toward the crossing. $t_{warning}$ was measured for different vehicle speeds. For each vehicle speed, we repeated measurement of $t_{warning}$ five times and obtained the average value of $t_{warning}$. At the same time,  the locations where an alert message was actually sent (pedestrian) and received (driver) were recorded based on an LED indicator and a camera to calculate $t_{warning}^{GT}$. Fig.~\ref{fig:driver_warning_vehicle_speed} shows the ground-truth distance to the crossing when the vehicle received an alert message, which varies due to the nondeterministic nature of the user walking speed, GPS locations, and message delay.

%\begin{figure}[!htbp]
%\centering
%\includegraphics[width=.9\columnwidth]{user_warning_threshold}
%\caption {Measured user warning time and collision probability compared with ground truth user warning time.}
%\label{fig:user_warning_threshold}
%\end{figure}

%\begin{figure}[!htbp]
%\begin{minipage}[b]{0.48\columnwidth}
%\centering
%\includegraphics[width=\columnwidth]{user_warning_threshold}
%\caption {User warning time and collision probability.}
%\label{fig:user_warning_threshold}
%\end{minipage}
%\hspace{1mm}
%\begin{minipage}[b]{0.48\columnwidth}
%\centering
%\includegraphics[width=\columnwidth]{driver_warning_vehicle_speed}
%\caption {Driver warning per vehicle speed.}
%\label{fig:driver_warning_vehicle_speed}
%\end{minipage}
%\end{figure}
Fig.~\ref{fig:user_warning_threshold} shows the user warning times for different vehicle speed and the corresponding probabilities of collision. The figure also shows the ground-truth user warning time. The results indicate that when the vehicle speed was high, the vehicle was closer to the crossing when the alert message was generated, leading to the small user warning time and greater collision probability. We compared the measured user warning time calculated based on $t_c$ and $t_p$ with the ground truth user warning time. The difference was between 0sec and 3sec, and the average difference was 1.6sec. Although the max difference of 3sec is a non-negligible amount of time considering the fast moving vehicle, it can be compensated by configuring the system to fire an alert message several seconds early.

%Theoretically, if we drive at 5mph, it would take about 31seconds to reach the crossing. In the mean time, a typical human can walk about 41m in 31seconds, which show the feasibility of low probability. On the other hand, if the driver maintains 20mph, it will take just about 8second to reach the crossing, and the pedestrian can move about 10meters. Considering that the user receives the warnings before reach the start of the crossing and the distance of the danger zone was 7m, we can see that there is indeed very high probability of collision.

%\subsubsection{Alerting the Driver}
%\label{sec:driver_warning}

%Besides, the collision probability would make sense only when the alert message was received by the driver. We measured the packet delivery rates by varying the vehicle speed. More specifically, we let the pedestrian continuously transmitted alert messages to the driver and measured the number of successfully received messages. In particular, to maintain the constant speed, we used the vehicle's cruise control system and used part of the vehicle trajectory as the acceleration zone to reach the desired speed (Figure~\ref{fig:test_site}). Figure~\ref{fig:speed_vs_pdr} displays the results which indicate that packets are in general very reliably received regardless of the vehicle speed although a slight decrease was observed for higher speed.

\section{Conclusion}
\label{sec:conclusion}

We have presented $\mathsf{SaferCross}$, a first fully functioning prototype mobile system for preventing distracted phone use. We develop critical system components for mobile systems for pedestrian safety focusing on the positioning accuracy, energy efficiency, activity detection, and effective risk assessment, laying the foundation for future research and development of mobile systems for pedestrian safety. We demonstrated that $\mathsf{SaferCross}$ effectively performs risk assessment of pedestrian safety via systematic integration of various software components for pedestrian positioning, phone use activity detection, energy efficiency, and car-to-pedestrian communication. We expect that the technical contributions made in this paper will be useful assets for various other transportation research involving pedestrians. A potential extension of this work is to enhance the pedestrian positioning accuracy as well as the proposed energy efficiency algorithm utilizing the recently arising 5G network{~\cite{dighriri2017comparison}}. Another interesting future direction is to understand how pedestrians and drivers respond to a warning sent by the proposed system, which is an important research problem as noted by a recent research{~\cite{rahimian2018harnessing}} that pedestrians tend to reduce their attention when receiving a warning, and they sometimes do not respond to a warning once they initiated a crossing.

\bibliographystyle{IEEEtran}
\bibliography{mybibfile}

\newpage

\begin{IEEEbiography}[{\includegraphics[width=1in,height=1.25in,clip,keepaspectratio]{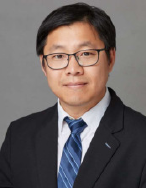}}]{Myounggyu Won} (M'13) received the Ph.D. degree in Computer Science from Texas A\&M University at College Station, in 2013. He is an Assistant Professor in the Department of Computer Science at the University of Memphis, Memphis, TN, United States. Prior to joining the University of Memphis, he was an Assistant Professor in the Department of Electrical Engineering and Computer Science at the South Dakota State University, Brookings, SD, United States from Aug. 2015 to Aug. 2018, and he was a postdoctoral researcher in the Department of Information and Communication Engineering at Daegu Gyeongbuk Institute of Science and Technology (DGIST), South Korea from July 2013 to July 2014.  His research interests include smart sensor systems, connected vehicles, mobile computing, wireless sensor networks, and intelligent transportation systems. He received the Graduate Research Excellence Award from the Department of Computer Science and Engineering at Texas A\&M University - College Station in 2012.
\end{IEEEbiography}

\begin{IEEEbiography}[{\includegraphics[width=1in,height=1.25in,clip]{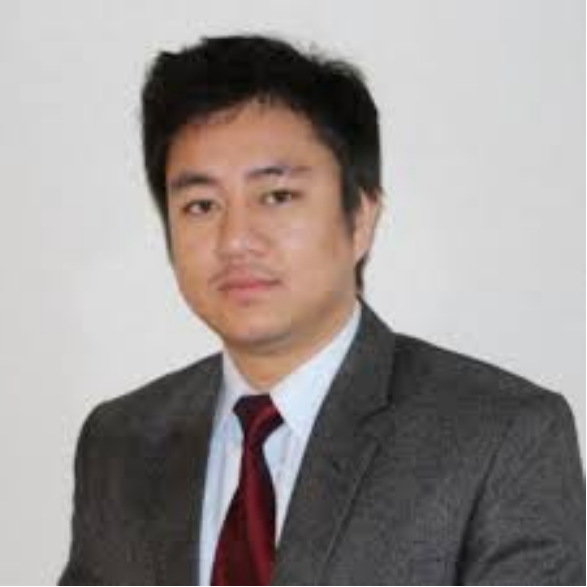}}]{Aawesh Shrestha} received the B.S. degree in the Computer Engineering Department from Kathmandu University, Kathmandu, Nepal in 2014. He also received the M.S. degree in the Department of Computer Science from the South Dakota State University, Brookings, United States in 2018. He is now working as an associate at the Deutsche Bank.
\end{IEEEbiography}

\begin{IEEEbiography}[{\includegraphics[width=1in,height=1.25in,clip,keepaspectratio]{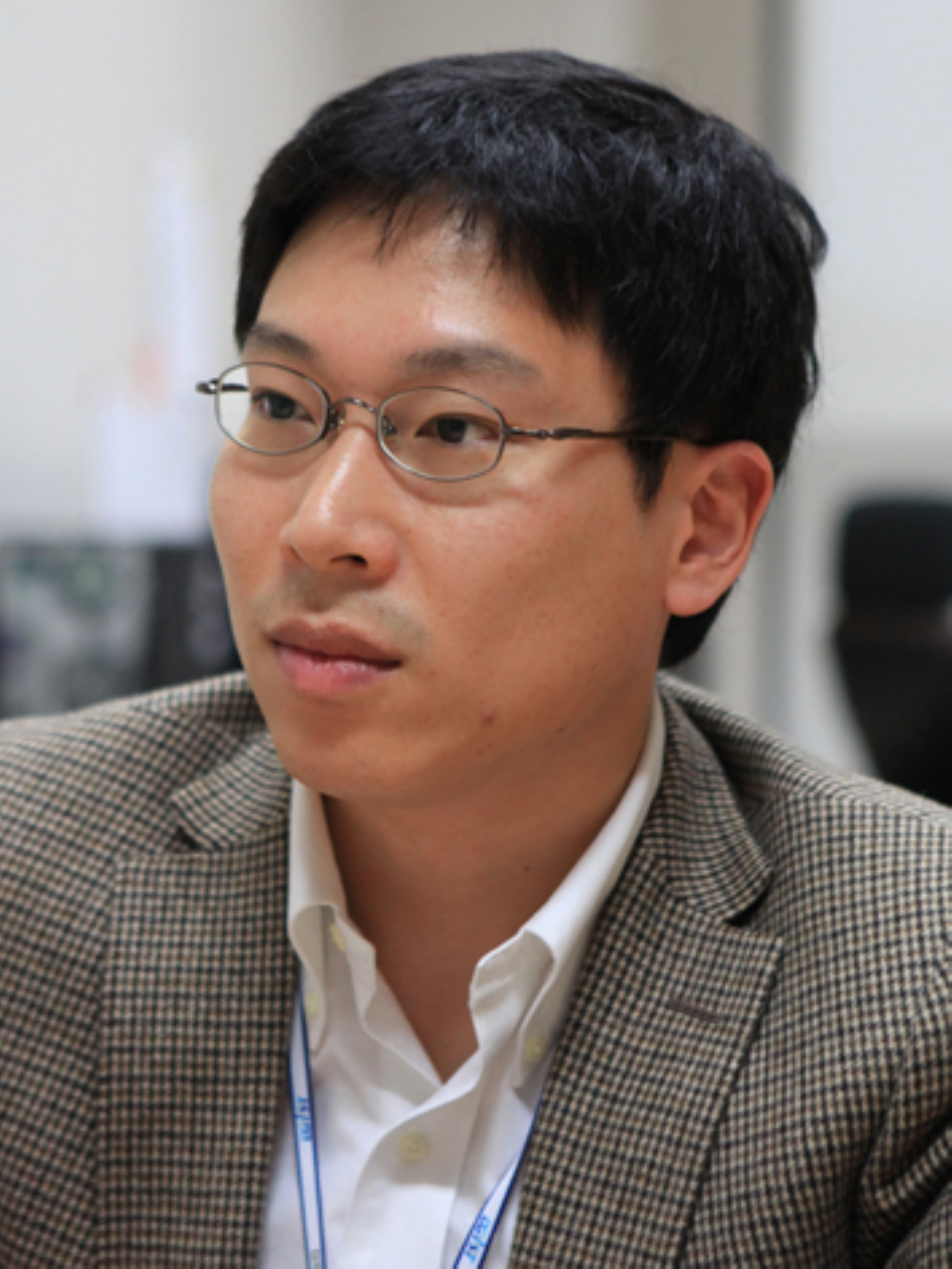}}]{Kyung-Joon Park} (M'05) received the B.S. and M.S. degrees in electrical engineering and the Ph.D. degree in electrical engineering and computer science from the School of Electrical Engineering, Seoul National University, Seoul, South Korea, in 1998, 2000, and 2005, respectively. From 2005 to 2006, he was a Senior Engineer with the Samsung Electronics, South Korea. From 2006 to 2010, he was a Postdoctoral Research Associate with the Department of Computer Science, University of Illinois at Urbana–Champaign, Champaign, IL, USA. He is currently a Professor with the Department of Information and Communication Engineering, Daegu Gyeongbuk Institute of Science and Technology, Daegu, South Korea. His research interests include resilient cyber-physical systems and smart factory.
\end{IEEEbiography}

\begin{IEEEbiography}[{\includegraphics[width=1in,height=1.25in,clip,keepaspectratio]{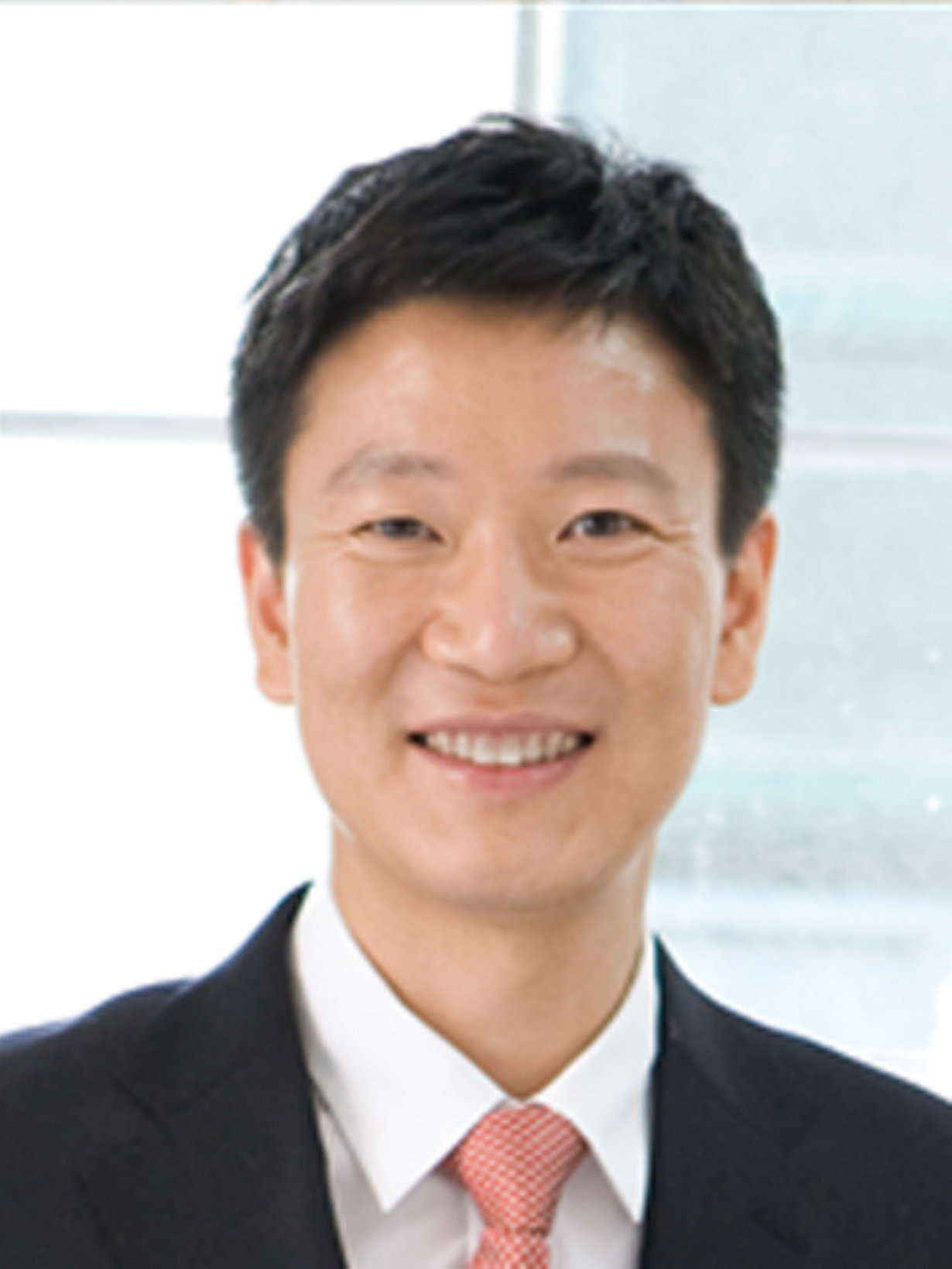}}]{Yongsoon Eun} (M'03--SM'19) received the B.A. degree in mathematics and the B.S. and M.S.E. degrees in control and instrumentation engineering from Seoul National University, Seoul, South Korea, in 1992, 1994, and 1997, respectively, and the Ph.D. degree in electrical engineering and computer science from the University of Michigan, Ann Arbor, MI, USA, in 2003. From 2003 to 2012, he was a Research Scientist with the Xerox Innovation Group, Webster, NY, USA, where he was involved in a number of subsystem technologies in the xerographic marking process and image registration method in production inkjet printers. He is currently a Professor with the Department of Information and Communication Engineering, Daegu Gyeongbuk Institute of Science and Technology, Daegu, South Korea. His research interests include control systems with nonlinear sensors and actuators, geometric control of quadrotors, communication networks, and resilient cyber-physical systems.
\end{IEEEbiography}

\EOD

\end{document}